\documentclass[%
aip,
 jmp,%
 amsmath,amssymb,
 reprint,%
]{revtex4-1}

\usepackage{graphicx}
\usepackage{dcolumn}
\usepackage{bm}

\begin{document}

\preprint{AIP/123-QED}

\title[Dynamic changes in network synchrony reveal resting-state functional networks]{Dynamic changes in network synchrony reveal resting-state functional networks}

\author{Vesna Vuksanovi\'{c}}
 \email{vesna.vuksanovic@bccn-berlin.de}
\author{Philipp H{\"o}vel}%
 
\affiliation{Institut f{\"u}r Theoretische Physik, Technische Universit\"at 
Berlin, Hardenbergstra\ss{}e 36, 10623 Berlin, Germany \\
}%
\affiliation{Bernstein Center for Computational Neuroscience, 
Humboldt-Universit{\"a}t zu Berlin, Philippstra{\ss}e 13, 10115 Berlin, Germany \\
}%

\date{\today}

\begin{abstract}
Experimental fMRI studies have shown that spontaneous brain activity i.e. in the absence of
any external input, exhibit complex spatial and temporal patterns of co-activity between segregated brain regions. These so-called
large-scale resting-state functional connectivity networks represent dynamically organized neural assemblies interacting
with each other in a complex way. It has been suggested that looking at the dynamical properties of complex patterns of
brain functional co-activity may reveal neural mechanisms underlying the dynamic changes in functional interactions.
Here, we examine how global network dynamics is shaped by different network configurations, derived from
realistic brain functional interactions. We focus on two main dynamics measures: synchrony and variations in synchrony.
Neural activity and the inferred hemodynamic response of the network nodes are simulated using system of 90
FitzHugh-Nagumo neural models subject to system noise and time-delayed interactions. These models are embedded into the
topology of the complex brain functional interactions, whose architecture is additionally reduced to its main structural
pathways. In the simulated functional networks, patterns of correlated regional activity 
clearly arise from dynamical properties that maximize synchrony and variations in synchrony. Our results on the
fast changes of the level of the network synchrony also show how flexible changes in the
large-scale network dynamics could be.\\ 
%
\end{abstract}

\keywords{ network dynamics; whole brain simulations; FitzHugh-Nagumo neural model}
\maketitle

\begin{quotation}
Experimental studies of the human brain activity at rest i.e. without any overt-directed behavior have revealed patterns of correlated activity, so called resting-state networks. The neural mechanisms contributing to the formation of these networks are largely unknown. We use modeling approach to interpret these experimental findings, looking at the brain as the dynamical system. We characterize brain network dynamical properties by synchrony and variability in synchrony. We demonstrate that functional brain interactions may arise from the network dynamics which allow flexible changes between different network configurations. We show that these changes reflect almost periodic alternations between network synchronized and desynchronized state.
\end{quotation}
\section{Introduction}
Well organized spatio-temporal low-frequency fluctuations ($<$ 0.1 Hz) 
have been observed in blood-oxygen-level-dependent (BOLD) functional magnetic resonance imaging (fMRI) signals of a 
mammalian brain in the absence of any stimulation or task-related behavior 
\cite{BIS95,DAM06,VIN07a}. The mechanisms generating these patterns of correlated activity, also called resting-state functional networks
are largely unknown. It has been suggested that they reflect a complex interplay between brain structural 
connections and the dynamics of interacting neuronal units 
\cite{BUL09,DEC11}. Theoretical models of the large-scale brain resting-state 
activity have explored the range of conditions that might govern intrinsic brain 
processes and contribute to the generation of correlated fluctuations between 
segregated brain areas \cite{HON07,HON09,CAB11}. Particular attention 
has been paid to the properties of the network dynamics that enable the emergence 
of a metastable state, which represents the network's tendency to 
switch between synchronized and desynchronized states and thus, explore different network configurations
\cite{GHO08a, DEC09,CAB11, ROY14}. One approach to observe metastable network states in brain dynamics is to consider
self-sustained oscillatory local neural dynamics and neural signal transmission delays  in the interactions between 
network elements \cite{CAB11,CAB12,DEC09}.  Similarly,  the coupling strength and signal transmission delays in the
interactions between noise-driven neural oscillators, enable generation of the spatio-temporal correlated low-frequency
($<$ 0.1 Hz) fluctuations in the dynamics of the resting brain \cite{GHO08a,GHO08}.

Other important ingredients of the large-scale brain dynamics models are complex brain network interactions expressed
usually in the form of the connectivity matrix. The matrix elements represent the anatomical connection strengths
inferred from all possible structural connections between anatomically defined regions of interest (ROIs). Using this
approach, the large-scale network dynamics models are built on matrices of the human \cite{HON09,CAB11,CAB12}  or monkey
\cite{HON07,GHO08, DEC09} brain architectures.  Both empirical and numerical findings suggest that the brain structural
and functional architectures (or networks) share many common features \cite{BUL09}. However, in contrast to 
structural networks, the functional networks of the brain undergo permanent reconfiguration of their connections depending on brain cognitive state \cite{BAS14,HEL14}. Moreover, altered topology of brain networks is indicator of pathological state \cite{CAB13b}.

We combine these approaches to investigate properties of the network dynamics that underlie patterns of spatio-temporal
correlations of resting-state functional networks. In addition, we consider the contribution of the long-distance
functional interactions i.e. those that are not supported by direct neural paths between the interacting brain areas. In
our previous work \cite{VUK14}, we have shown that the long-distance functional correlations may emerge from
relay-like interactions between neural oscillators that share large parts of their individual network neighborhoods.
Here, we aim to explore, if these types of the spatio-temporal relations may also emerge in the network dynamics
subject to system noise and signal propagation delays. For this purpose, we chose to model the local node dynamics by
noise-driven excitable FitzHugh-Nagumo (FHN) neurons oscillating at 15 Hz. Our approach is similar to those of Ghosh and
colleagues \cite{GHO08}, however, different in the two main model ingredients: (i) we use a
connectivity matrix derived from diffusion weighted brain imaging data, and (ii) we take into account functionally
realistic network interactions that are also supported by physical (structural) connections. Additionally, in our
analysis of the simulated dynamics, we focus on network metastability rather than linear approximation of its
stability.

Through simulations of computational models of resting-state functional correlations, we show that the neural
network dynamics display fast transitions between synchronized and desynchronized states. These transitions give rise to
spatio-temporal correlation patterns that resemble to those found in resting-state fMRI experiments.

\begin{figure*}[!ht]
\begin{center}
{\bf(A)}\includegraphics[width=0.4\textwidth]{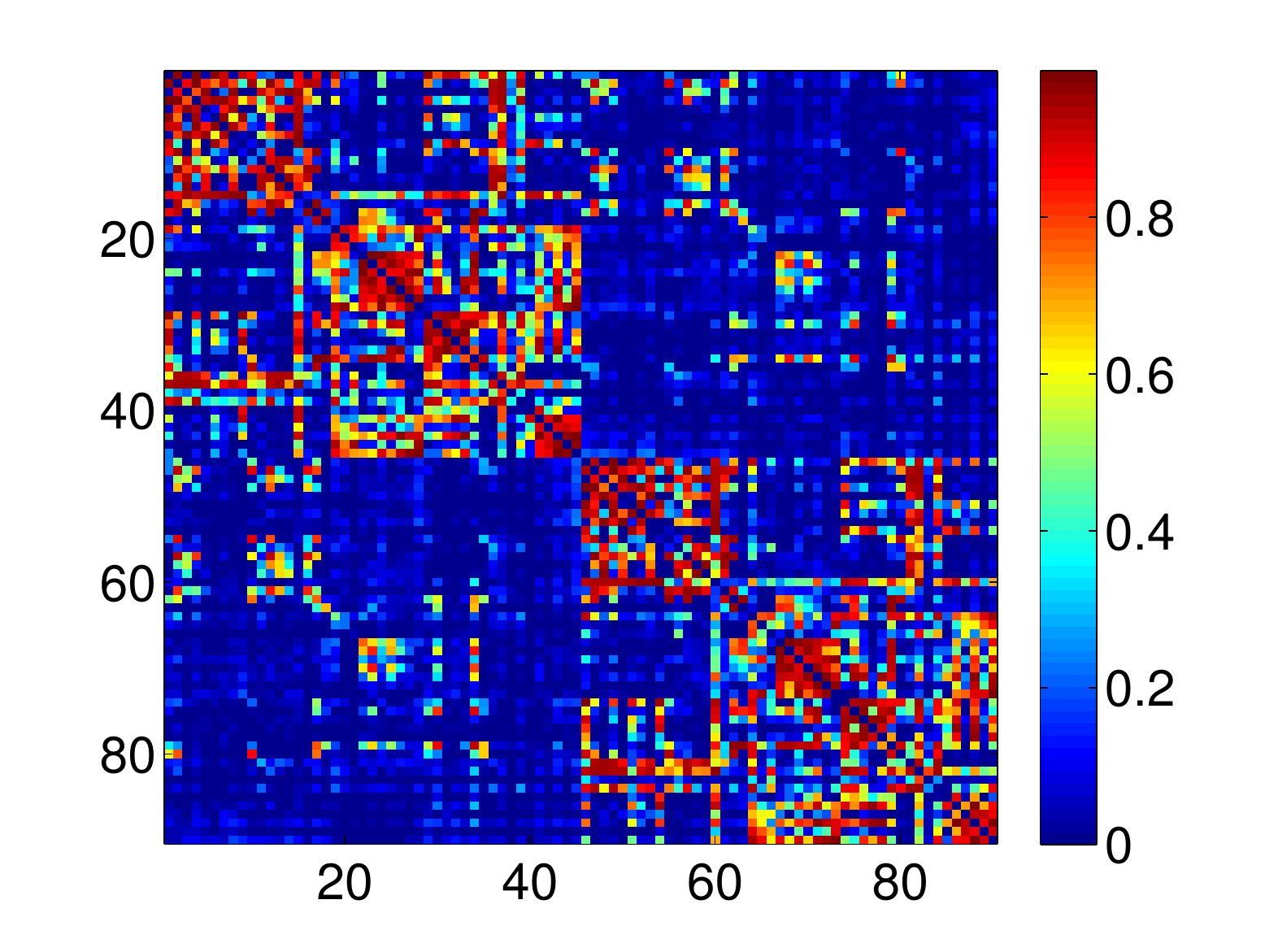}
{\bf(B)}\includegraphics[width=0.4\textwidth]{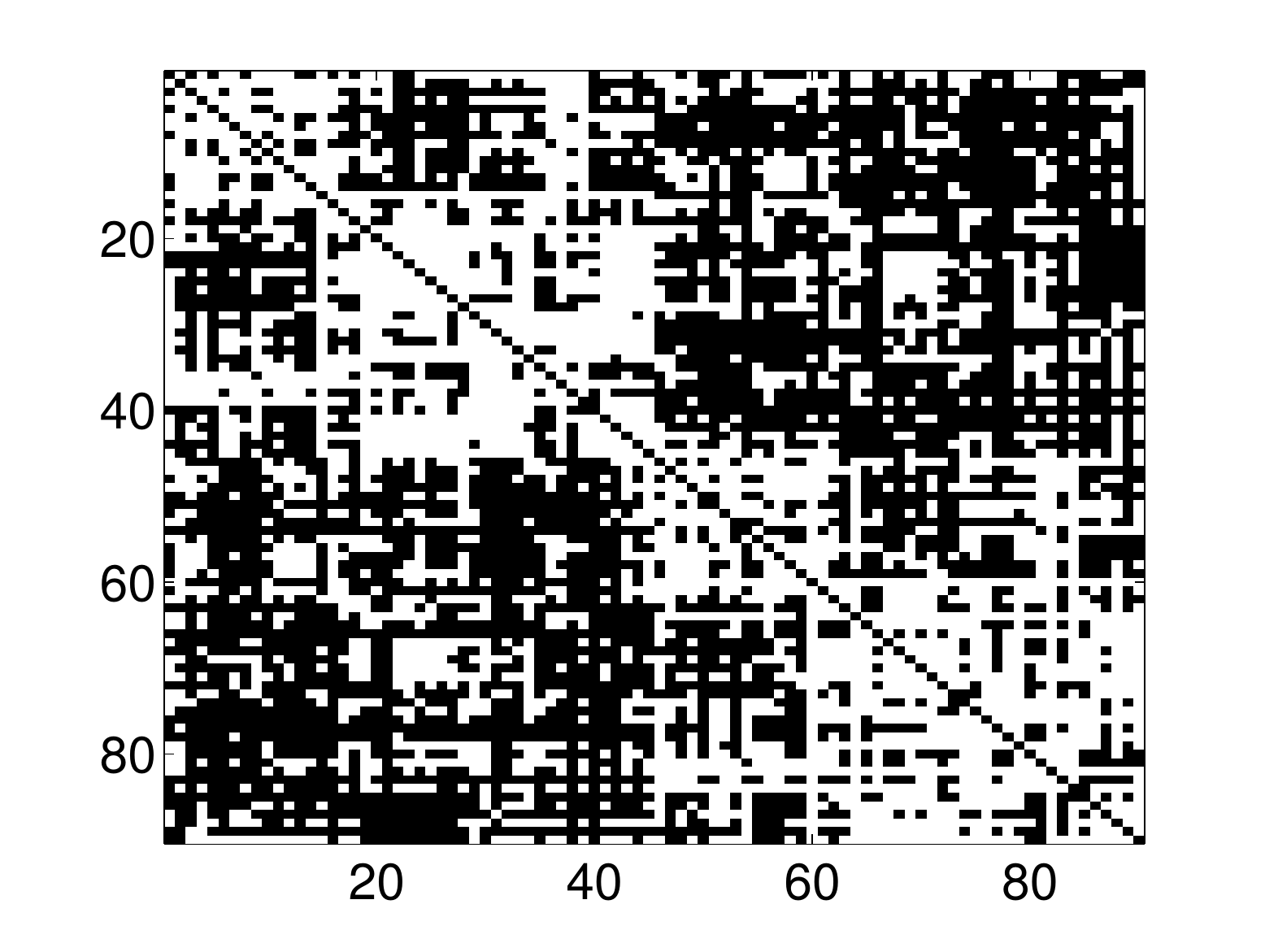}\\
{\bf(C)}\includegraphics[width=0.4\textwidth]{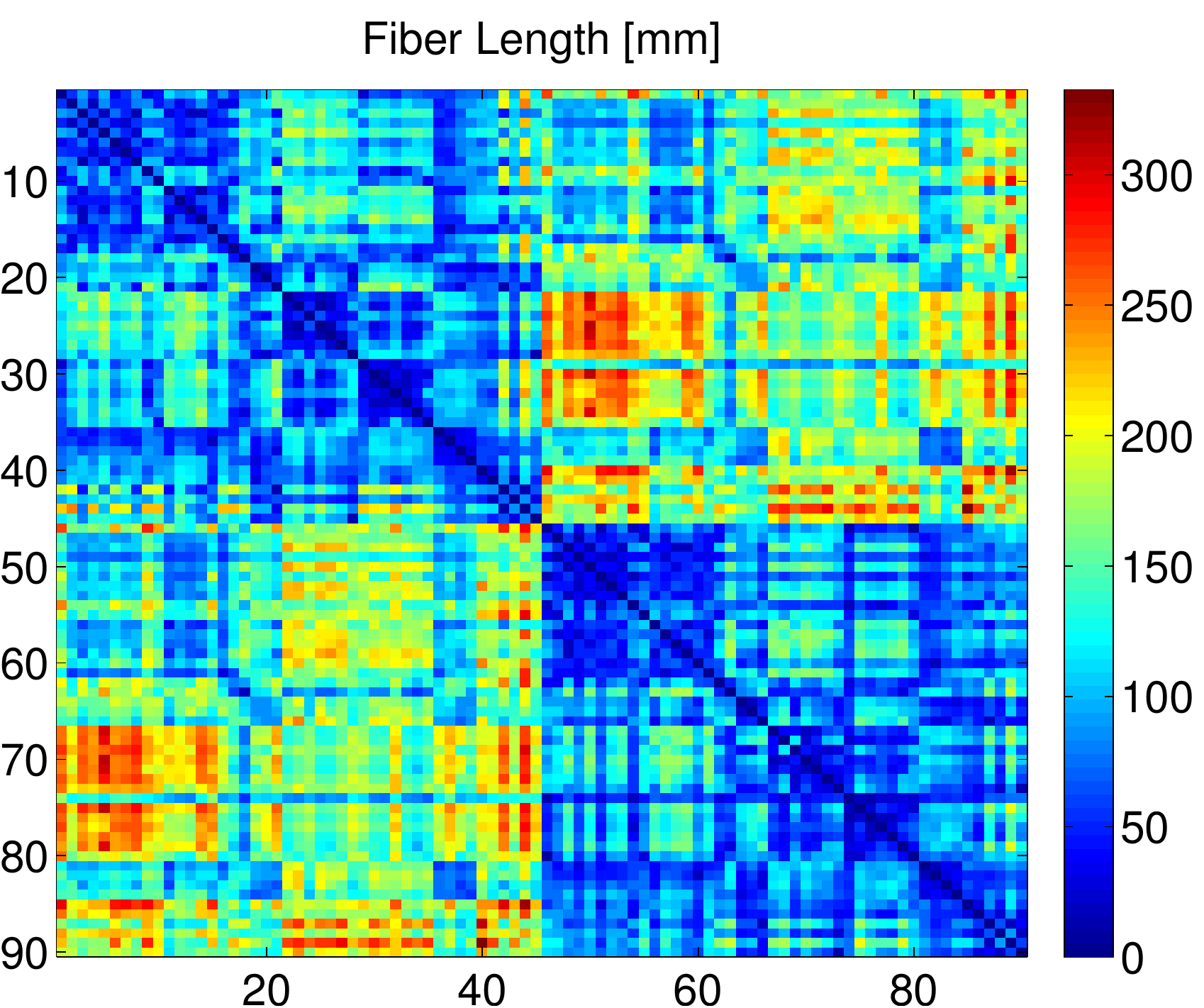}
{\bf(D)}\includegraphics[width=0.4\textwidth]{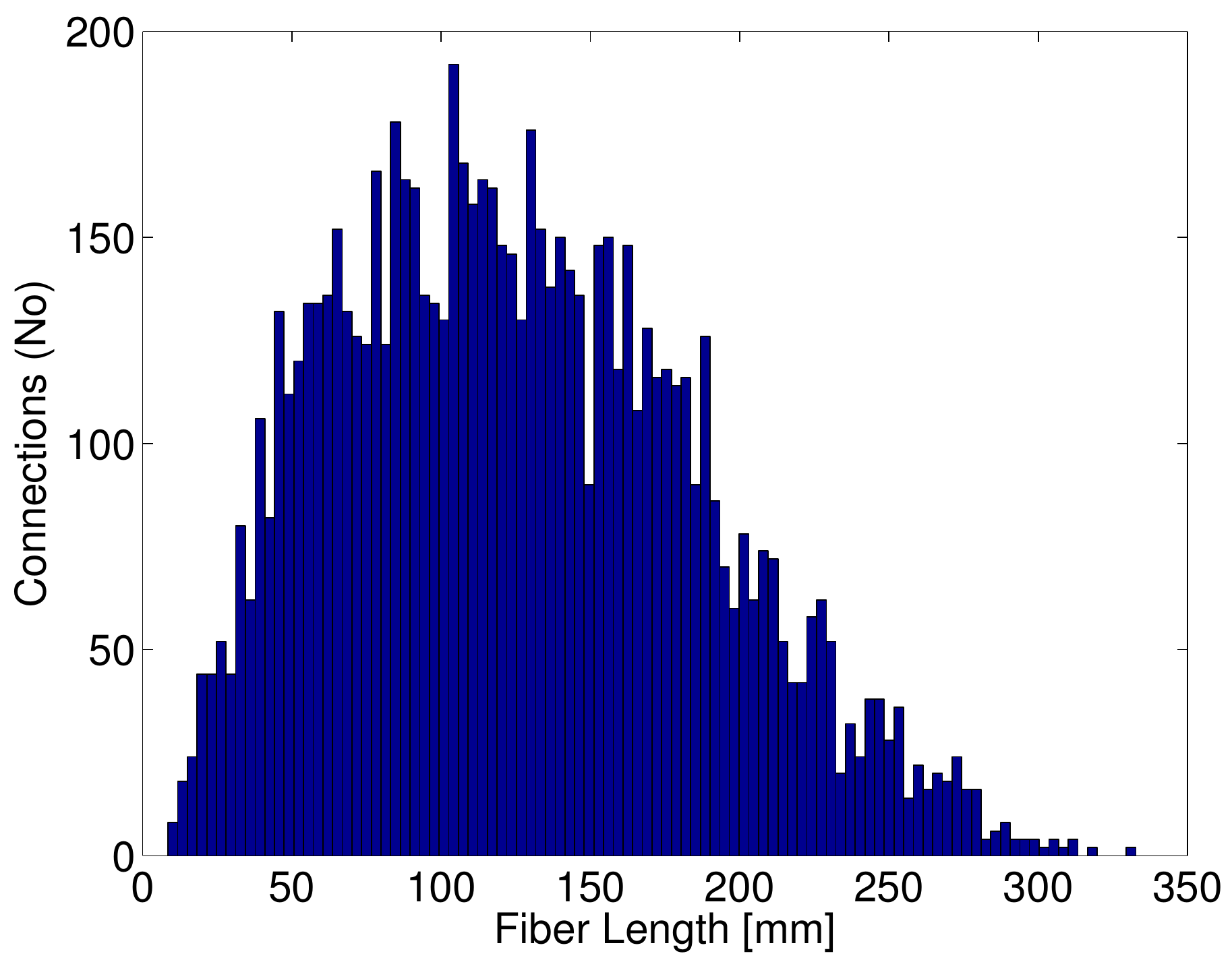}
\end{center}
\caption{
{\bf Functionally realistic connectivity between 90 brain regions considered in the 
model.} {\bf(A)} Anatomical connectivity probability (ACP) matrix and {\bf(B)} its binarized version when the threshold
$p = 0.05$ is applied. Application of the threshold to the ACP matrix enables also that the homologous interactions,
which are usually not good resolved by DW-DTI analysis method, are represented with the equal probabilities in the
model. {\bf(C)} The fiber lengths between 
the pairs of nodes and {\bf(D)} corresponding number of the connections. 
The matrices are ordered according to the regions listed in Table~\ref{tab:AAL}. First 45 regions belong to the right
hemisphere. 
}
\label{acp_fc}
\end{figure*}
\section{Methods}
\subsection{Extraction of network topology}
To build a model of the neural network dynamics on functionally realistic network interactions, we have combined
empirically derived structural and resting-state functional maps similarly to our previous study \cite{VUK14}. In short,
these maps were extracted from the 90 brain regions parcellated according to the Tzourio-Mazoyer brain atlas
\cite{TZO02} using the automated 
anatomical 
labeling (AAL) method. The full list of regions considered in the analysis 
is given in the Table~\ref{tab:AAL}. 

The structural connectivity is estimated from diffusion weighted brain imaging data, according to the procedure
described in \cite{ITU08}. The procedure maps out probabilities for the presence of the direct neural connections
between any pairs of the 90 considered anatomical ROIs. We will therefore refer to this map as to the anatomical
connection probability (ACP) map or matrix \cite{ITU08}. The functional connectivity (FC) matrix is extracted from
BOLD fMRI data of 26 subjects \cite{VUK14, MAR07}. The matrix entries represent pairwise
temporal correlation coefficients between ROIs mean BOLD time series estimated from the total scanning time (7.5
minutes). 

\begin{figure}
\begin{center}
\includegraphics[width=0.5\textwidth]{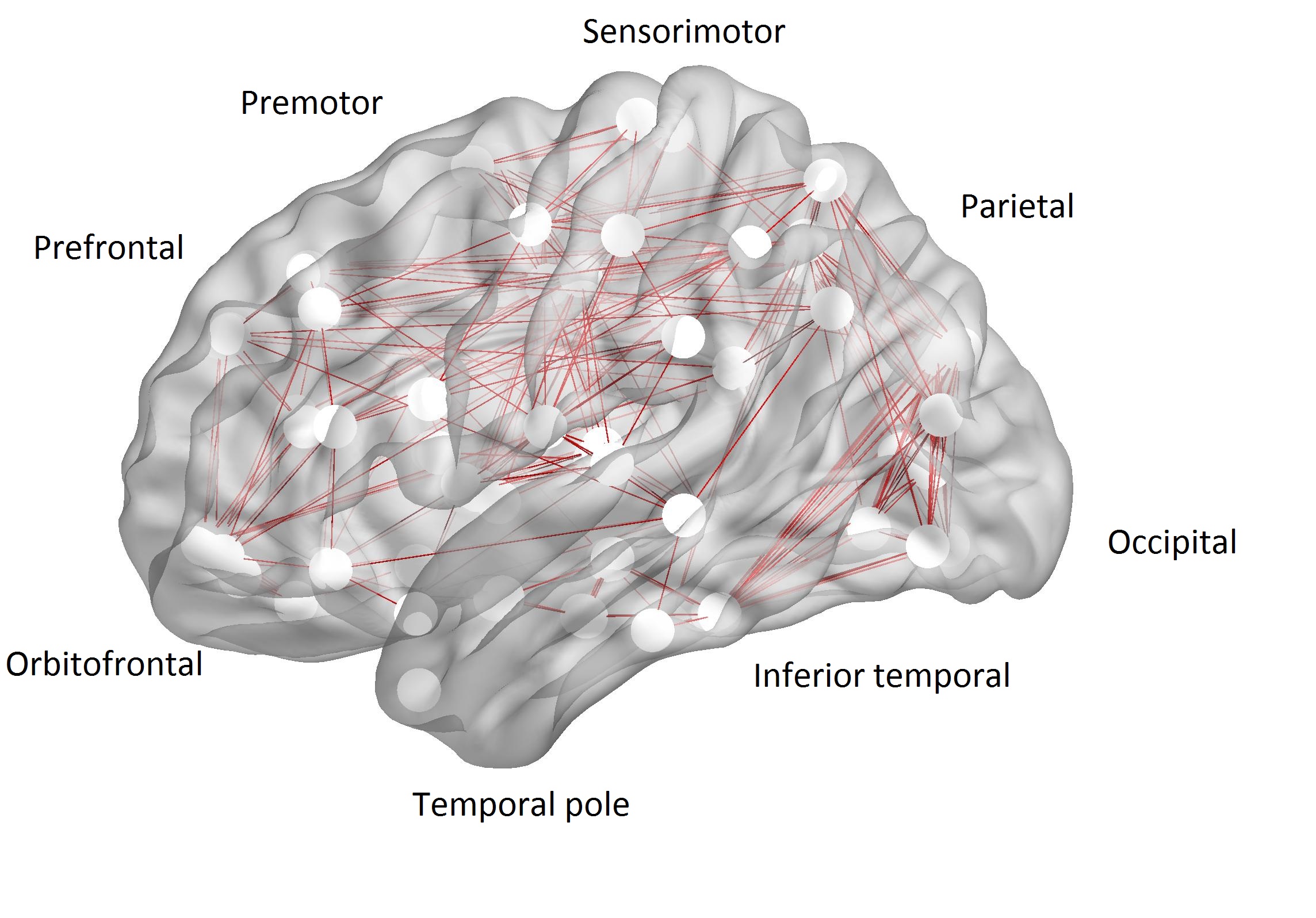}
\end{center}
\caption{
{\bf Visualization of brain functional networks.} The position of the network 
nodes is given according to the coordinates of the centre of the mass of the 
AAL cortical regions in MNI space. Size of the node is proportional to its 
degree for threshold $r = 0.57$. List of regions is given in Table~\ref{tab:AAL}.
}
\label{brain}
\end{figure}

To derive the coupling topology for the network model used in the simulations, we combined structural and functional
connectivity matrices in the following way. Given the statistical nature of FC and ACP entries we first applied a
binarization method to extract only statistically significant connections. By doing that we also took into account the
conditions that allow comparisons between brain structural and functional networks \cite{WIJ10}. Binarization is done by
applying thresholds $r$ and $p$ for the FC and ACP matrix, respectively. If matrix values are greater than or equal to
the threshold values, then the corresponding element of the adjacency matrix is set to 1; otherwise it is set to 0. For
the FC matrix, we have used thresholds in the interval $r = [0.52, 0.53, \dots, 0.65]$, and for the ACP matrix
only one value $p = 0.05$ (see Fig.~\ref{acp_fc}). The lowest value for $r$ and the value for $p$ are chosen such that 
both networks have equal connection density $\kappa = 0.5$ \cite{WIJ10}. The upper boundary 
for $r$ is the highest threshold at which the FC 
network still comprises one component. We then use the element-wise product of these binarized matrices as a coupling
topology for the simulations. As a result, we preserve those 
strongly pronounced interactions of the functional networks that are connected via direct anatomical 
links, that is, regions without direct anatomical links are not directly coupled in 
the simulations. This procedure is previously described in detail in \cite{VUK14}. For an exemplary visualization of
the resulting brain network see Fig.~\ref{brain}.
\subsection{Simulation of network dynamics - neural and BOLD activity}
To infer the BOLD signal in dependence on the network properties, we first 
simulate the underlying neural activity. We consider the neural dynamics on the 
network of $N = 90$ cortical regions whose 
local dynamics are represented by the homogeneous 
FitzHugh-Nagumo (FHN) neurons. 

\subsubsection{Network of FitzHugh-Nagumo neural models}
\begin{figure}[!ht]
\begin{center}
{\bf(A)}\includegraphics[width=0.4\textwidth]{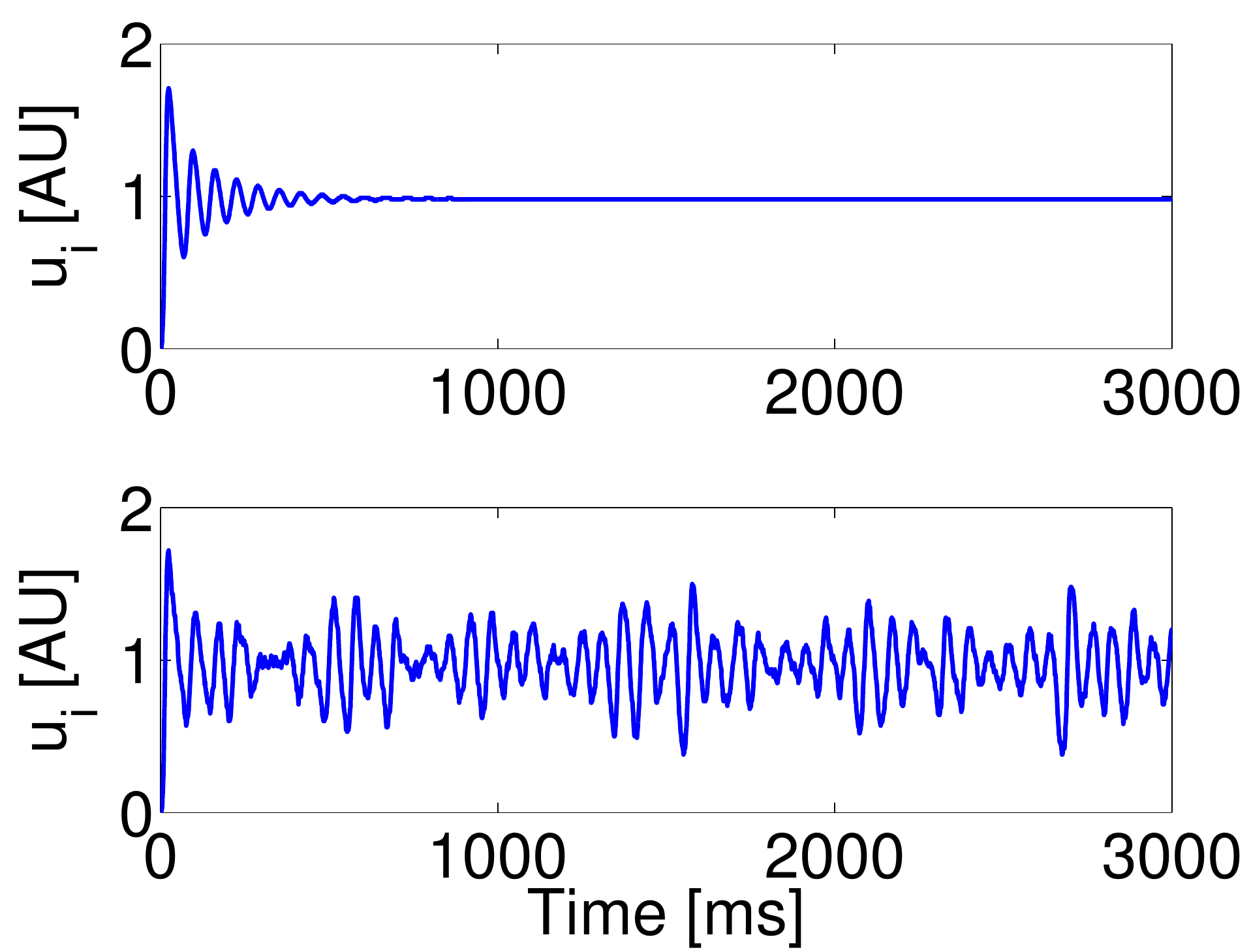}
{\bf(B)}\includegraphics[width=0.4\textwidth]{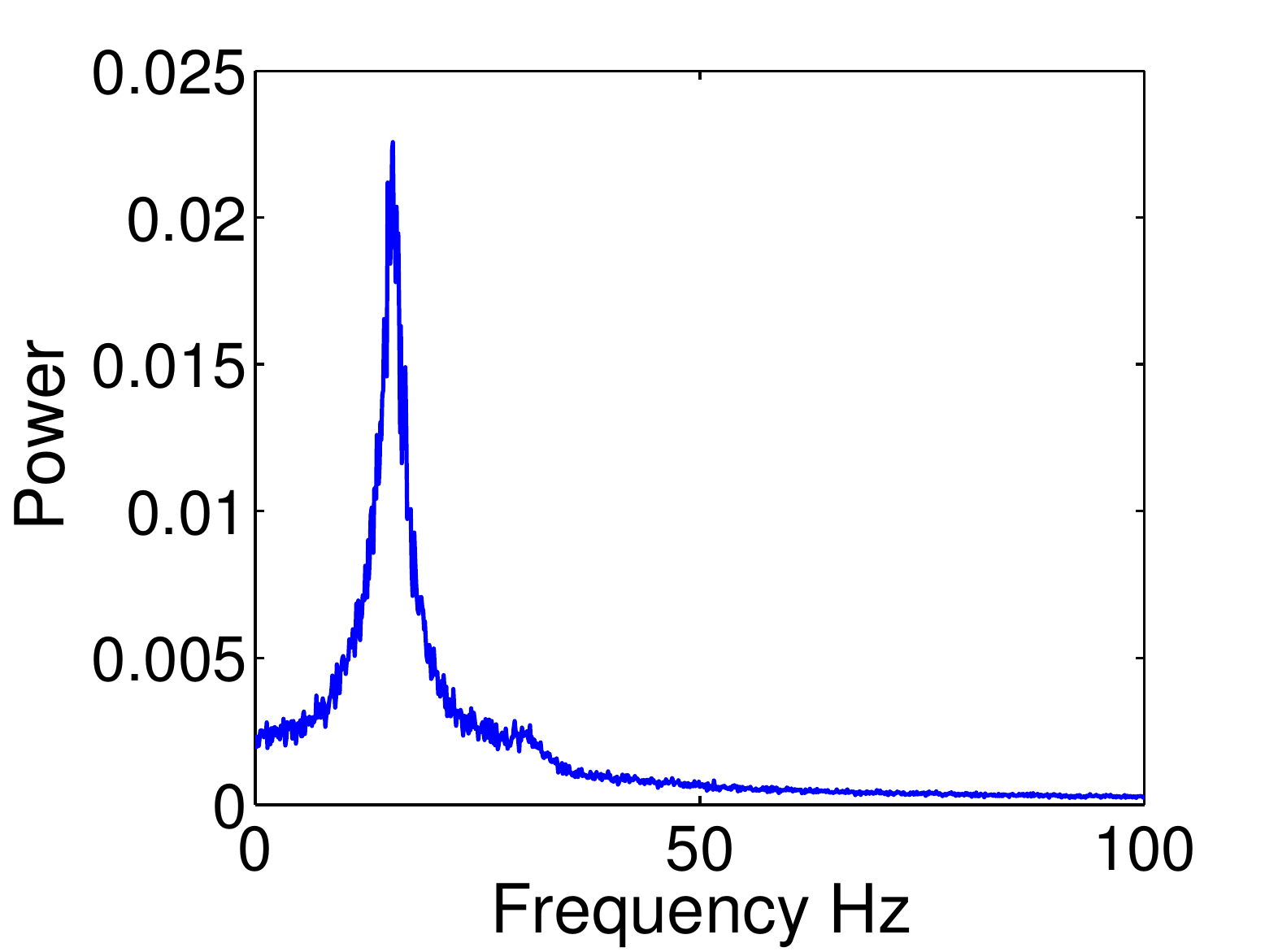}\\
\end{center}
\caption{
{\bf Time evolution of the activator $u_{i}$ variable of the FitzHugh-Nagumo 
neural model for an isolated node.} {\bf(A)} Top panel represents node dynamics 
without noise and bottom panel shows dynamics of the same node when noise is 
added via the term $n_{u}$, as described in the Eq.~\ref{eq:FHN_nw}. {\bf(B)} Fourier 
spectrum of the neural activity of an isolated node under the system noise. The peak frequency is at around 15 Hz.
System parameters: $\alpha = 0.85$, $\beta = 0.2$, $\gamma = 1.0$, $\tau = 1.25$, and $p=0.05$.
}
\label{node_ts}
\end{figure}

We model neural network dynamics by embedding individual FHN neural model into each 
of the 90 cortical regions, according to the connectivity matrix $\left\{a_{ij}\right\}, i,j= 1, \dots,N$.
FHN neural model can be described by two state 
variables $u$ and $v$, representing activator (membrane potential) and 
inhibitor (recovery variable), according to the following dynamical equations: 
\begin{subequations}
\begin{align} 
  \dot{u} &= g(u,v) = \tau \left(v + \gamma u - \frac{u^3}{3}\right)\\
  \dot{v} &= h(u,v) = -\frac{1}{\tau}\left(u - \alpha + \beta v -I\right),
\end{align}
\end{subequations}
where $I$ is magnitude of an external stimulus, which is assumed to be 
0 \cite{GHO08}. In order to obtain the dynamics of an isolated node in a damped oscillatory regime [as shown in
Fig.~\ref{node_ts} (A -- upper panel)], we consider the following system parameters throughout this paper:  
$\alpha = 0.85$, $\beta = 0.2$, $\gamma = 1.0$, and $\tau = 1.25$ .

To combine FHN units in a network, we utilize the following equations \cite{GHO08}: 
\begin{subequations}
\begin{align} 
   \dot{u}_i &= g(u_i,v_i) - c {\sum_{j=1}^N a_{ij}u_j(t - \Delta t_{ij})} + 
n_u\\
   \dot{v}_i &= h(u_i,v_i) + n_v,
\end{align}
\label{eq:FHN_nw}
\end{subequations}
where $u_i$ and $v_i$ are the activator and inhibitor variables, located at node $i$. $c$ denotes a global coupling
parameter ($c > 0$) and $a_{ij}$ are the elements of the connectivity matrix, defined above. $\Delta t_{ij}$ denote 
time delays. $n_u$ and $n_v$ are two independent additive white Gaussian noise 
terms with zero mean, unity variance and noise strength $D$. The influence of the
applied noise level on dynamics of an isolated node is shown in the 
Fig.~\ref{node_ts} (A -- lower panel). When coupled in the network, the dynamical behavior of a node changes to the
oscillatory regime. The frequency distribution of the noise-driven oscillations of a node is centered around 15 Hz as
shown in Fig.~\ref{node_ts}(B).

We solve the system of coupled differential 
equations with time delays and 
additive noise using the \textsc{Python}-module \textsc{pydelay} \cite{FLU09a}. 
The algorithm is based on the Bogacki-Shampine method \cite{BOG89,SHA01a}, which 
is also implemented in \textsc{Matlab's dde23} routine. We calculate time delays for a 
physiologically realistic value of the signal propagation velocity $v$ via 
$\Delta t_{ij} = d_{ij}/v$, where $d_{ij}$ are given by the lengths of the 
actual three-dimensional trajectory of the fiber tract between centers 
of the regions, i.e. network nodes $i$ and $j$. A color-coded representation of 
the $d_{ij}$ values is shown in Fig.~\ref{acp_fc}(C). We use a single value for the signal propagation velocity $v = 7 $
m/s
throughout the study.

\begin{figure*}[!ht]
\begin{center}
\hspace{0.5cm} \includegraphics[width=0.45\textwidth]{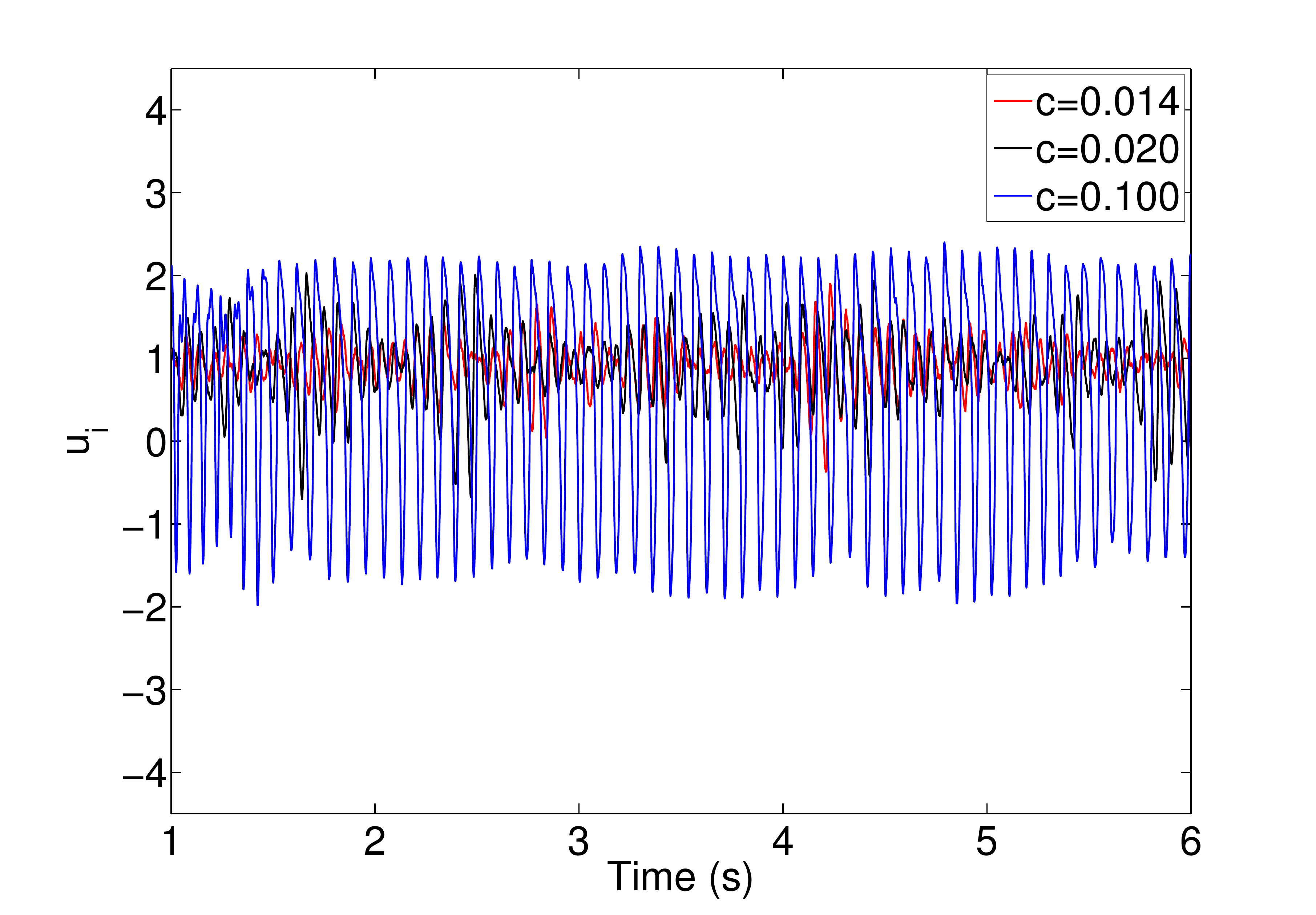}
\hspace{0.5cm}\includegraphics[width=0.45\textwidth]{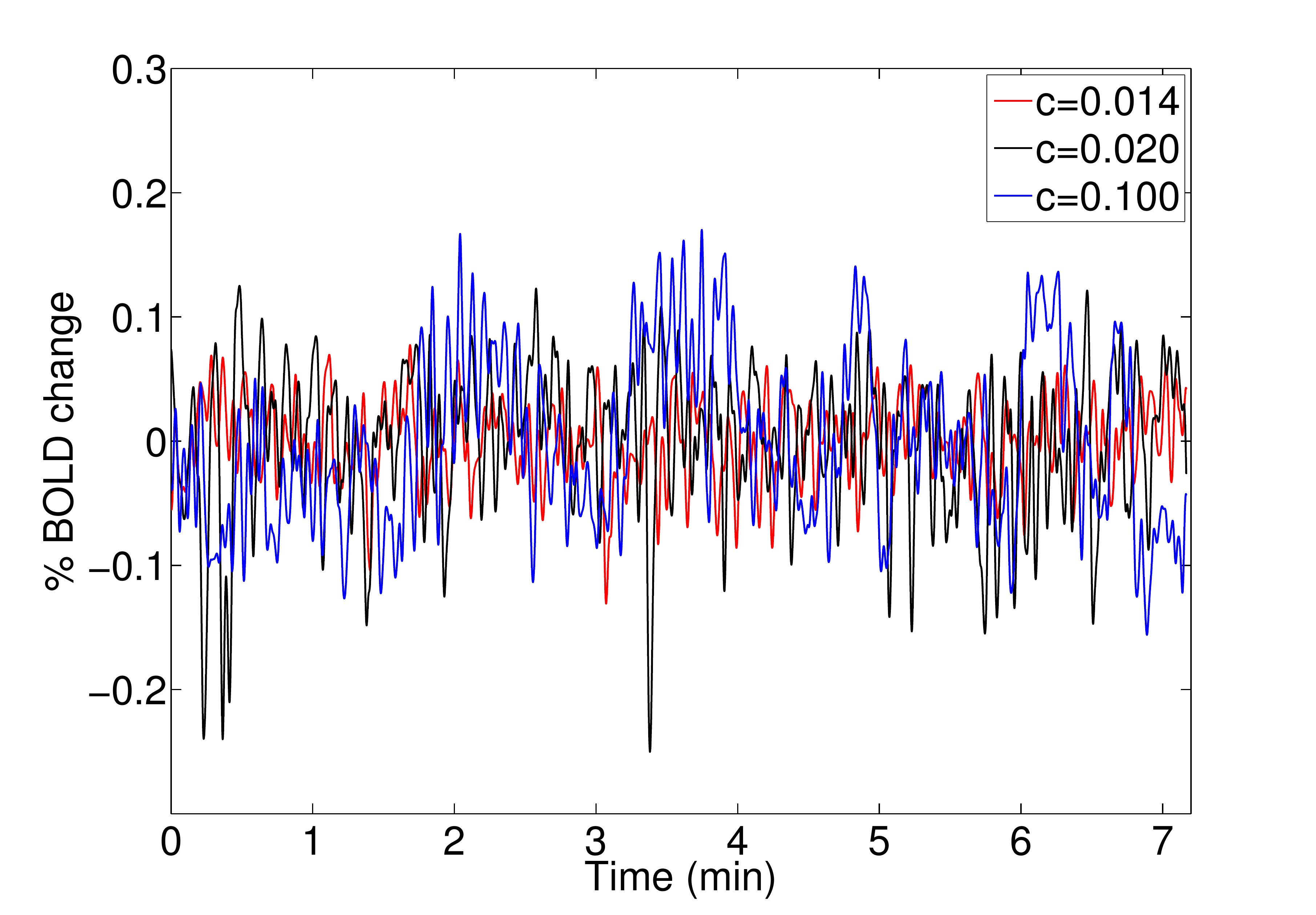}\\
{\bf(A)}\includegraphics[width=0.45\textwidth]{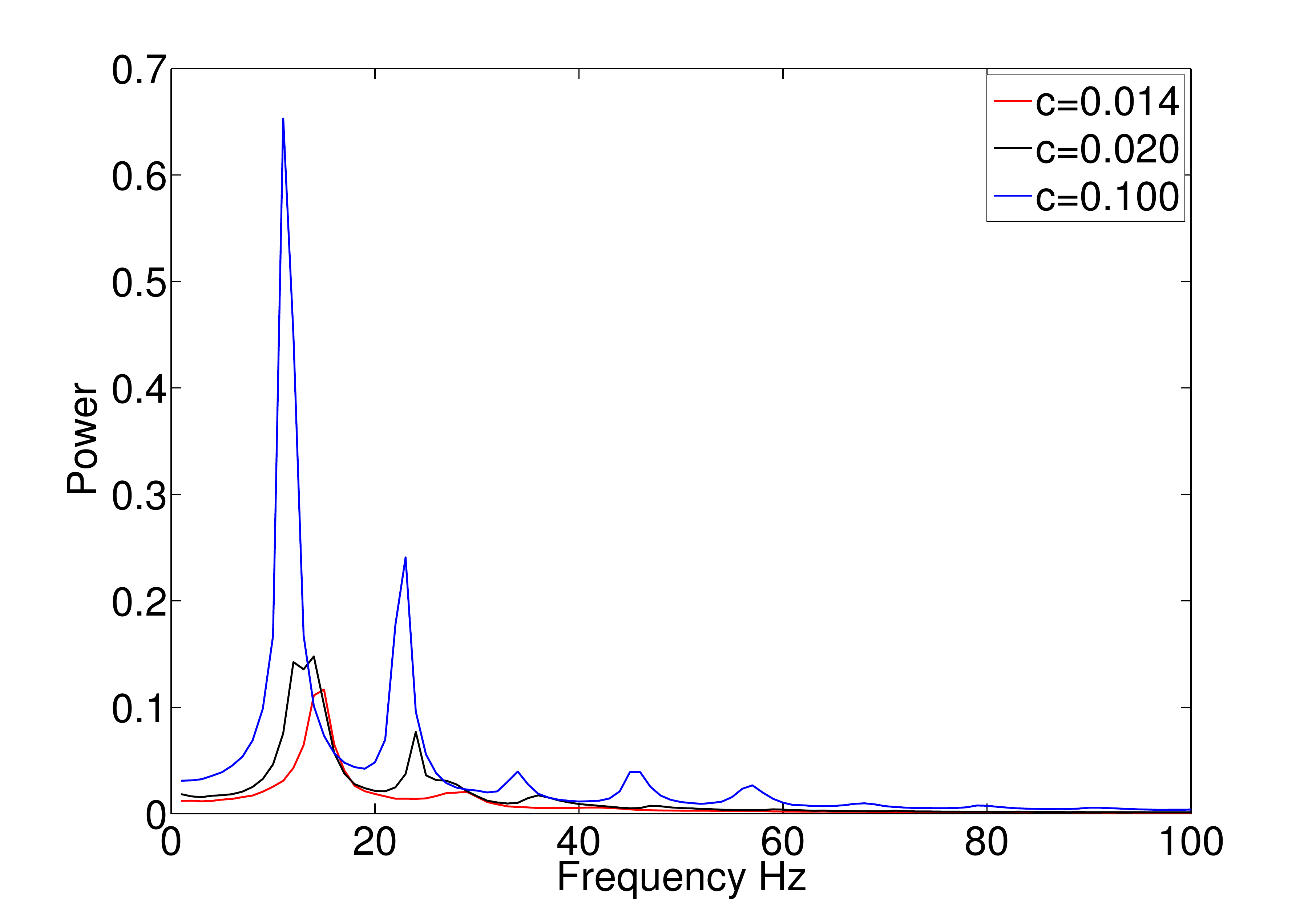}
{\bf(B)}\includegraphics[width=0.45\textwidth]{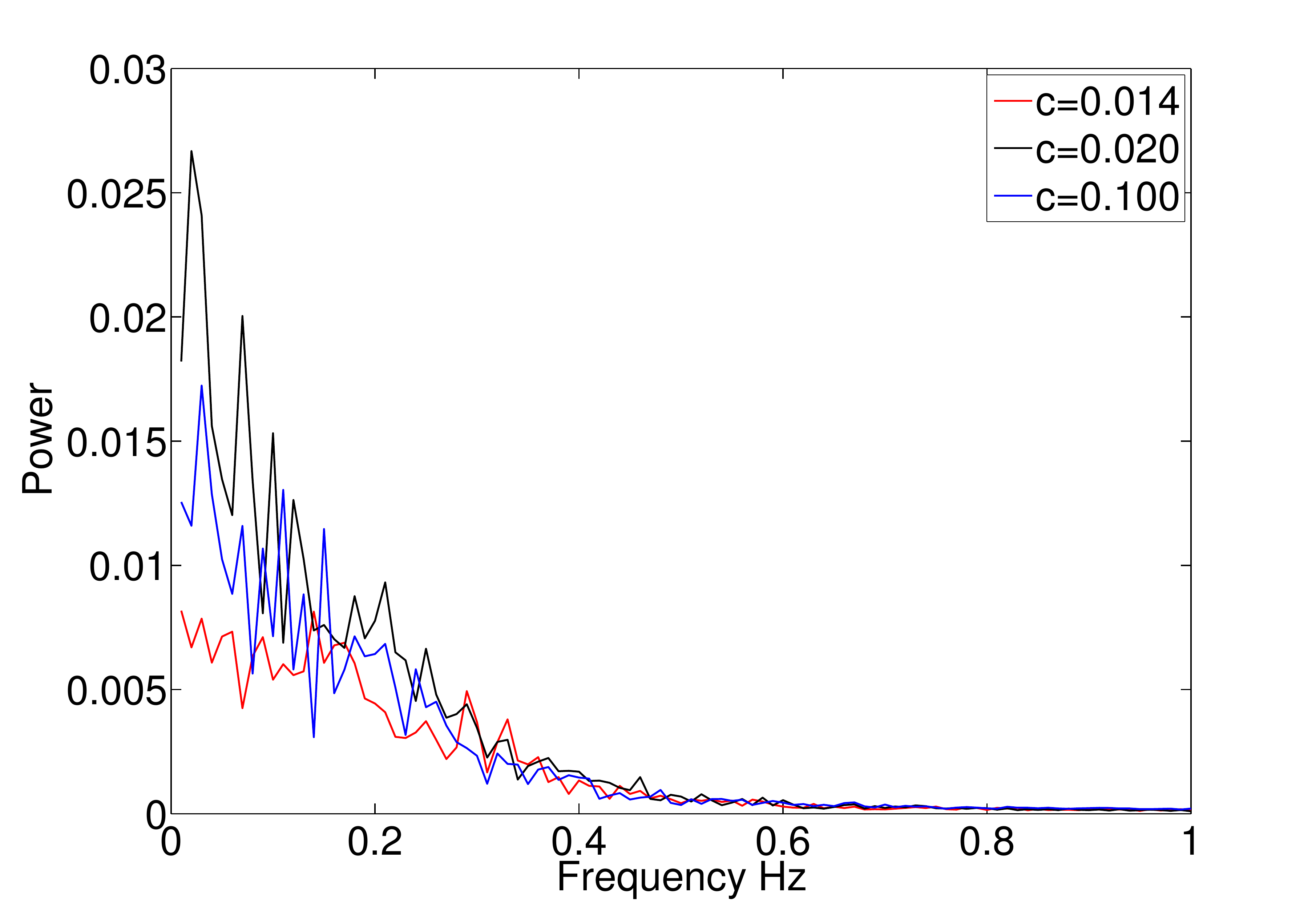}\\
\end{center}
\caption{
{\bf Time series and the corresponding Fourier power spectra} of {\bf(A)} neural and {\bf(B)} BOLD activity of the 
 representative network node ($i = 28$) for three different coupling strengths $c$ and a fixed threshold $r = 0.57$. The
node has 20 nearest neighbors for this particular threshold, which corresponds to the average number of the links in the
network. System parameters as in Fig.~\ref{node_ts}.} 
\label{fft_node}
\end{figure*}

We simulated 450\,s of the real-time neural 
activity sampled at a time step of 0.001\,s 
for a range of the coupling strengths $c$ and thresholds $r$. 
When time delays, system noise and global coupling term are taken into account, 
the simulated dynamics of the neural activity exhibits a behavior exemplary
shown in Fig.~\ref{fft_node}.
\subsubsection{Measures of the network dynamics}

To quantify global network dynamics we define two different measures: the synchrony of the neural activity and
metastability, which is defined by fluctuations in synchrony over time. Here, we have followed a general framework to
study synchronization phenomena in a network of coupled oscillators, given by Kuramoto and colleagues
\cite{STR00,IZH06b}. To apply this framework on the FHN system we first need to define phase-like variables $\theta_i$.
This can be done using one of the several approaches valid for the FHN models \cite{NGU11,OME13}. However, due to
the presence of the noise term in our model, we have chosen to use the Hilbert transform. The transform gives a time
series of complex numbers with a real and an imaginary part, which then can be used to calculate instantaneous phases 
$\theta_i$ for the each node as the argument of this complex number \cite{PIK01}. 

The synchrony of the simulated network dynamics is evaluated across all oscillators in the network, using the order parameter $R(t)$ \cite{STR00}:
  \begin{equation}
    R(t) = \left| \left\langle e^{i\theta_j(t)} \right\rangle \right|, \quad j=1,\dots,N,
    \label{eq:eq4}
  \end{equation}
where $\langle \cdot \rangle$ denotes average over all nodes in the network. \\
The order parameter measures the cooperative dynamics in the network and can have any value between 0 
and 1 \cite{NIC13}. The closer
$R(t)$ to unity, the higher the level of global synchrony in the network. 
The network is in a desynchronized state for $R(t) = 0$. 
    
In neural networks, however, 
$R(t)$ is never equal to 1 nor 0, i.e. neural networks never reach a fully 
synchronized nor desynchronized state. Instead, the brain network dynamics exhibits large variability,
and the amount of global synchrony of network nodes vary over 
time, indicating transitions from a synchronized to more desynchronized state \cite{DEC12,DEC12a}. 
This metastable state in brain dynamics can be quantified by the
standard deviation $\sigma_R$ of the order parameter $R(t)$ \cite{CAB11,SHA10}. It has been suggested that the brain
network dynamics operate in a state that maximizes synchrony and metastability \cite{SHA10, HEL14}.\\
\begin{figure*}[!ht]
\begin{center}
{\bf(A)}\includegraphics[width=0.45\textwidth]{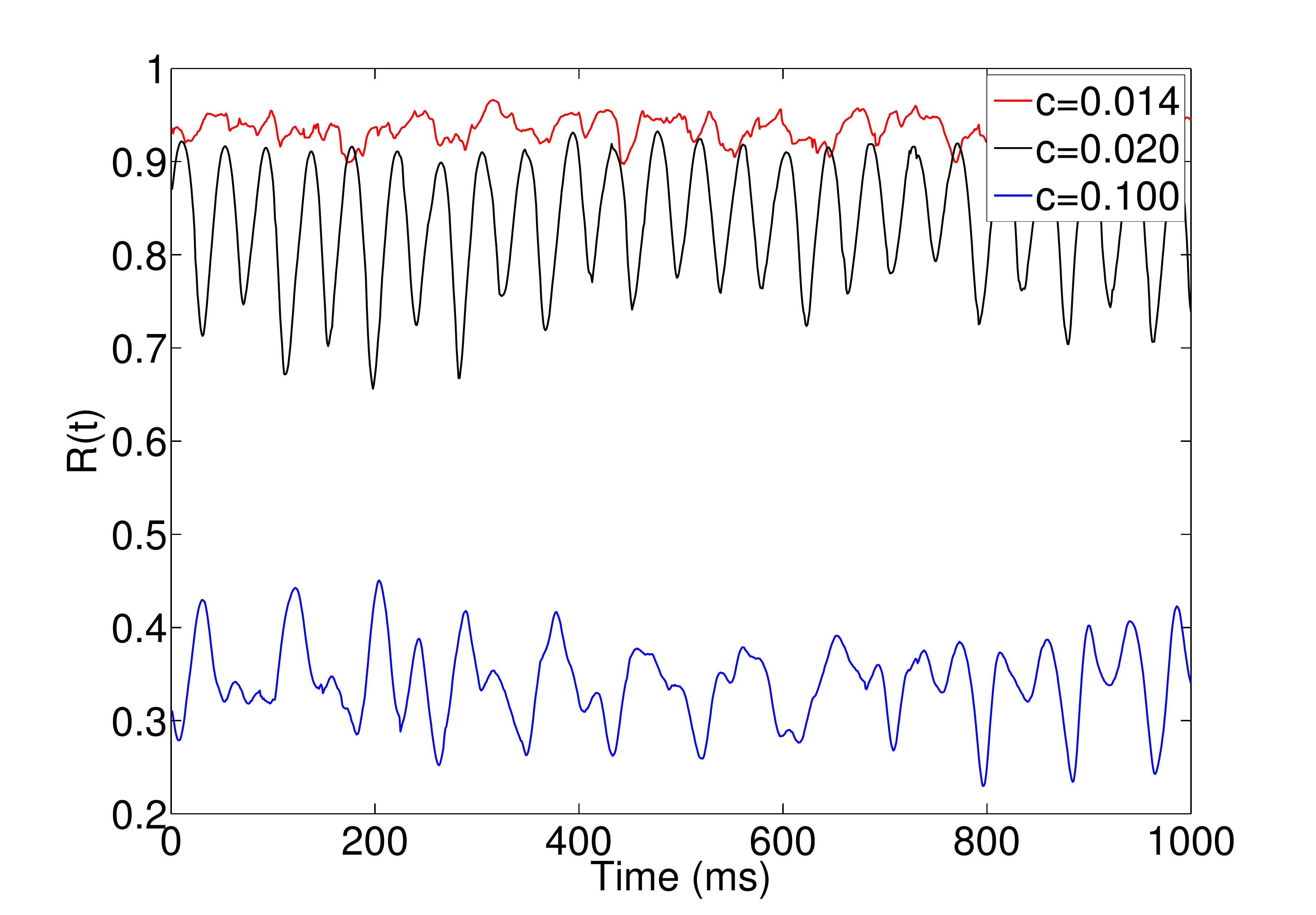}
{\bf(B)}\includegraphics[width=0.45\textwidth]{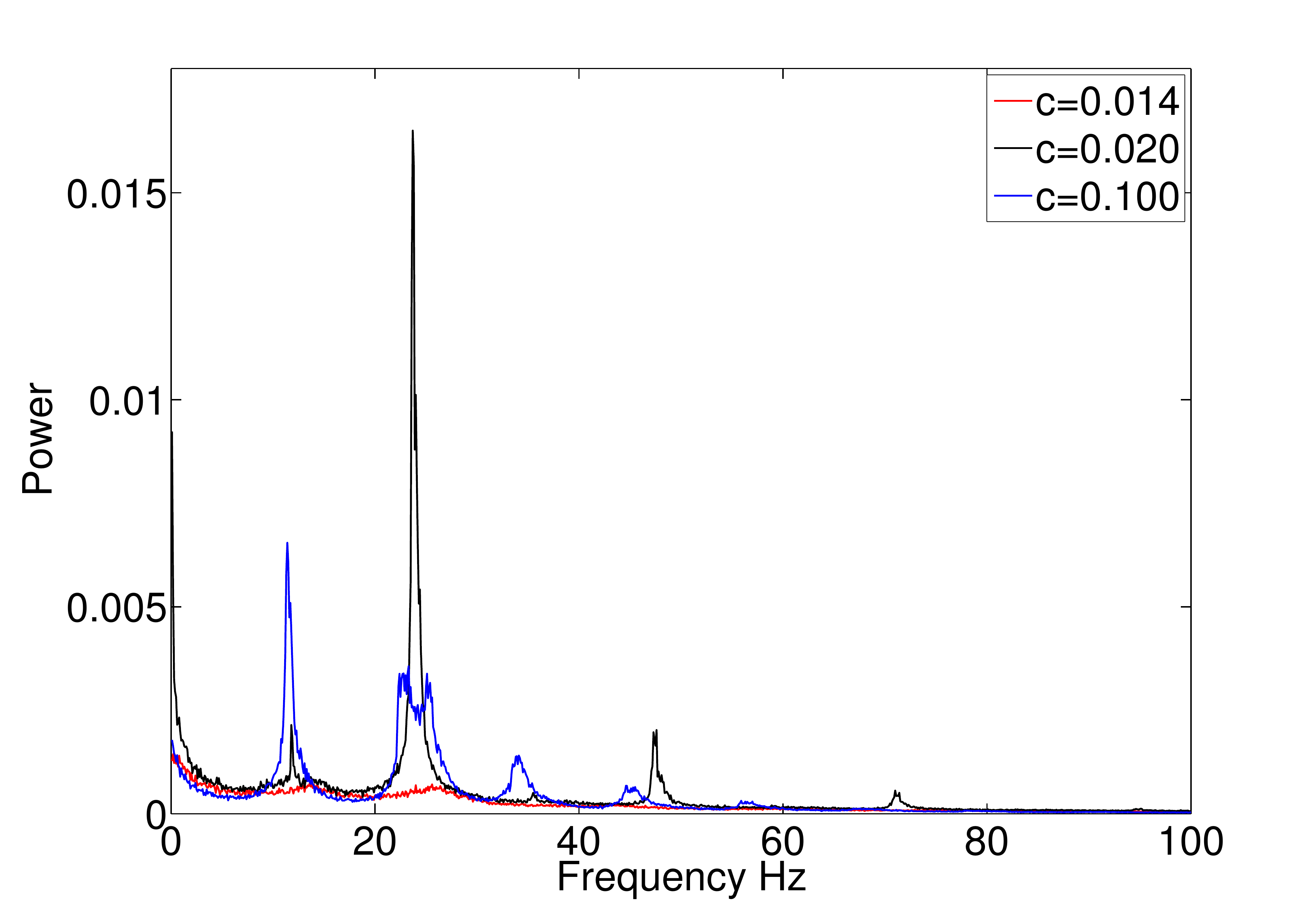}
\end{center}
\caption{
{\bf Global synchrony} as a function of {\bf(A)} time and {\bf(B)} frequency, for three different coupling strengths $c$
as indicated in the legend, and the network topology at the threshold $r = 0.57$. System parameters as in
Fig.~\ref{node_ts}.}
\label{fft_sync}
\end{figure*}
\subsubsection{Modeling BOLD activity and functional networks}
Low-frequency ($< 0.1$ Hz) oscillatory dynamics of BOLD signal is inferred from 
the simulated neural activity using the Balloon-Windkessel model 
\cite{FRI00}. The model describes the relationship between 
changes in the regional blood flow, caused by local neural activity, and the 
BOLD signal, which is presented as a function of 
venous volume and deoxyhemoglobin content. We implemented the model using 
default parameters values provided within SPM8 \cite{FRI94}. 

An important component of the BOLD model is a neural signal as a main input. It 
can be given in the form of either neural spiking rate or local field potential 
\cite{SET12}. To calculate BOLD activity, we feed the time series of the activator 
variables $u_i$, which resemble membrane potentials, into the model. 
Representative time series of modelled BOLD activity and corresponding 
power spectra are shown in Fig.~\ref{fft_node}(B) for different coupling strengths.

In order to quantify the functional connectivity between nodes $i$ and $j$ from 
the simulated BOLD activity, we calculate pairwise Pearson correlation 
coefficient:
\begin{equation} 
  \mathrm\rho(i,j) = \frac{\langle V_i(t)V_j(t)\rangle - \langle 
V_i(t)\rangle 
\langle 
V_j(t)\rangle}{\sigma\left(V_i(t)\right) \sigma\left(V_j(t)\right)},
\end{equation}
where $V_i(t)$ represents the time series of the activity of the node $i$ at 
time 
$t$, $\sigma$ is its standard deviation, and $\langle \cdot \rangle$ denotes 
temporal averages. 

\subsubsection{Complex network measures of simulated functional networks}
In our simulations, we use the network interactions -- derived from experimental fMRI and DW-DTI data -- as coupling
topologies. They can be characterized using graph-theoretical measures across range of 
correlation thresholds. In the Fig.~\ref{network_meas} we show some measures derived from the coupling topology at the matrix threshold $r = 0.57$. The following network properties are calculated 
as defined in Ref.~\cite{RUB10}:

\textit{Degree:} The number of connections with the other nodes in the network.
Degree of a node $i$ is calculated
\begin{equation}
   k_i = \sum_{j=1}^N a_{ij},
\end{equation}
where, $a_{ij}$ represents connection between nodes $i$ and $j$: $a_{ij} = 1$ 
if there is a connection between nodes and $a_{ij} = 0$ otherwise.

\textit{Clustering coefficient:} Ratio of the number of connections in the 
neighborhood 
of a node and the number of connections if the neighborhood was fully 
connected. Clustering coefficient of the network is calculated as
\begin{equation}
   C = \frac{1}{N} \sum_{i=1}^N 
C_i = \frac{1}{N} \sum_{i=1}^N \frac{2t_i}{k_i(k_i)-1},
\end{equation}
where $C_i$ is the clustering coefficient of node $i$ and $t_i$ is number of 
triangles around a node $i$.

\textit{Global efficiency:} The average inverse shortest path length between 
the node 
and its neighbors. Global efficiency of the network is calculated as
\begin{equation}
   E = \frac{1}{N}\sum_{i=1}^N 
E_i=\frac{1}{N}\sum_{i=1}^N \frac{\sum_{j=1,j\neq i} 
^N d_{ij}^{-1}}{N-1},
\end{equation}
where the shortest path length $d_{ij}$ between nodes $i$ and $j$ is defined 
by the number of links between the respective nodes, that is,
\begin{equation}
   d_{ij} = \sum_{a_{uv}\in g_{i\leftrightarrow j}} a_{uv},
\end{equation}
where $g_{i\leftrightarrow j}$ is the shortest path between $i$ and $j$. Note 
that $d_{ij} = \infty$ for all unconnected pairs $i$, $j$.

\textit{Characteristic path} of the network:
\begin{equation}
   L = \frac{1}{N}\sum_{i= 1}^N
L_i=\frac{1}{N}\sum_{i= 1}^N \frac{\sum^N_{{j= 1},j \neq i} {d_{ij}}}
{N-1},
\end{equation}
where $d_{ij}$ is the shortest path length (distance) between nodes $i$ and 
$j$ defined above.

\begin{figure*}[!ht]
\begin{center}
{\bf(A)}\includegraphics[width=0.45\textwidth]{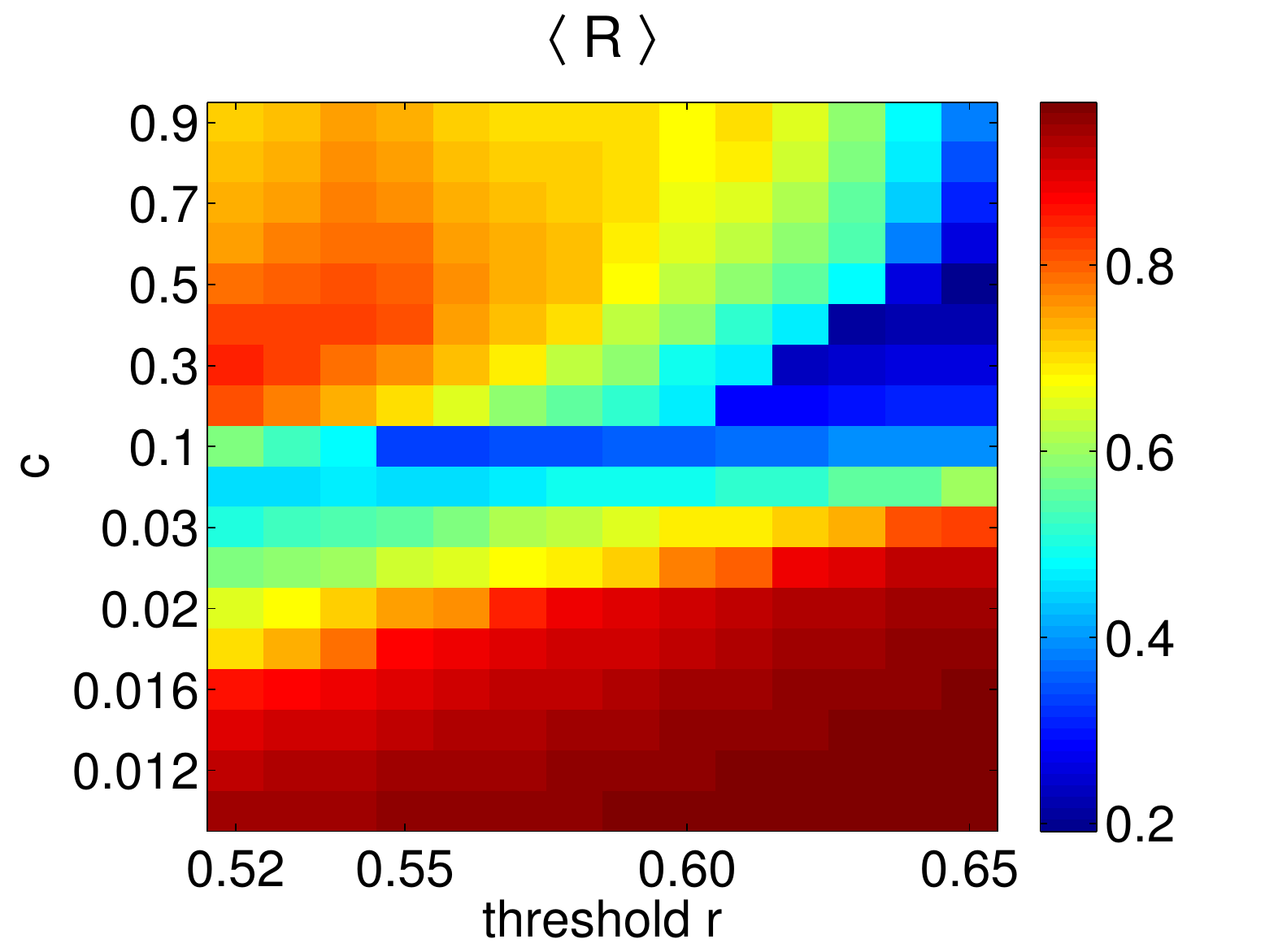}
{\bf(B)}\includegraphics[width=0.45\textwidth]{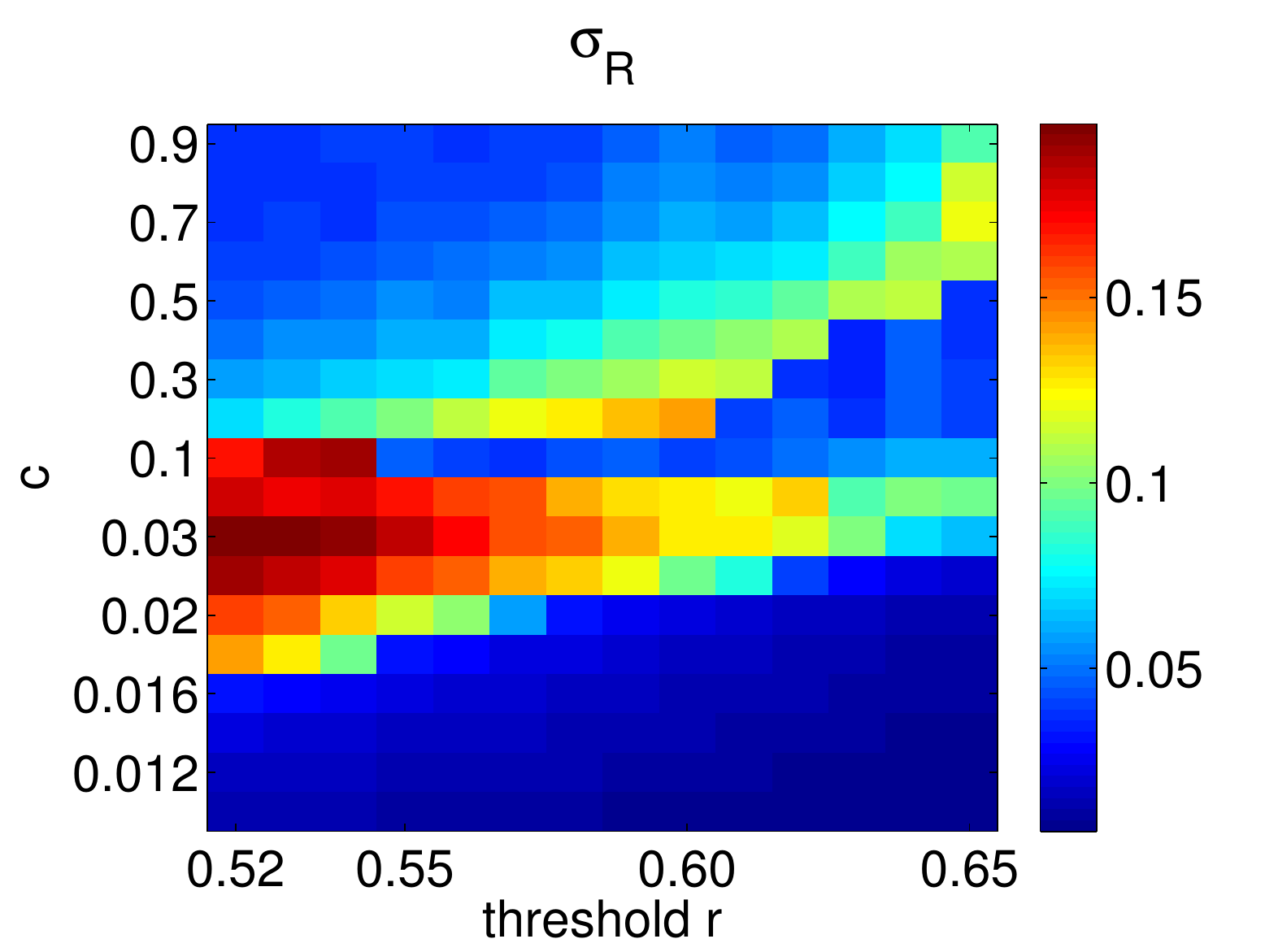}
\end{center}
\caption{
{\bf Global measures of the network dynamics.} {\bf(A)} Mean order parameter $\langle R(t)\rangle$ and {\bf(B)}
metastability, i.e. variations in synchrony $\sigma_R$ in the parameter space of the global coupling strength $c$ and 
the correlation threshold $r$. System parameters as in Fig.~\ref{node_ts}.}
\label{sync}
\end{figure*}

\section{Results}
In our simulations we consider functionally realistic network interactions, i.e. those that take into account
structurally supported connectivity between functionally related network nodes, and chose to model each node's dynamics
by a simple FHN neuronal model. The node dynamics is tuned such that each isolated node displays damped oscillations
(see Fig.~\ref{node_ts}(A) upper panel). However, noise added to the system drives each FHN
element from its equilibrium to an oscillatory state. The frequency of the oscillations of an isolated node with the
added noise is around 15 Hz (as shown in Fig.~\ref{node_ts}). We aim to explore how the global network dynamics is
influenced by the underlying topology and the strengths of the interactions. Therefore, focusing on the role of the
network interactions in the collective behavior, we kept neural signal transmission velocity at
the constant value ($v = 7 $ m/s). This represents a biologically realistic value for the neural signal propagation and
also an 
optimal value to obtain desired brain network dynamics in the system of the FHN elements \cite{GHO08}.

\begin{figure*}[!ht]
\begin{center}
{\bf(A)}\includegraphics[width=0.45\textwidth]{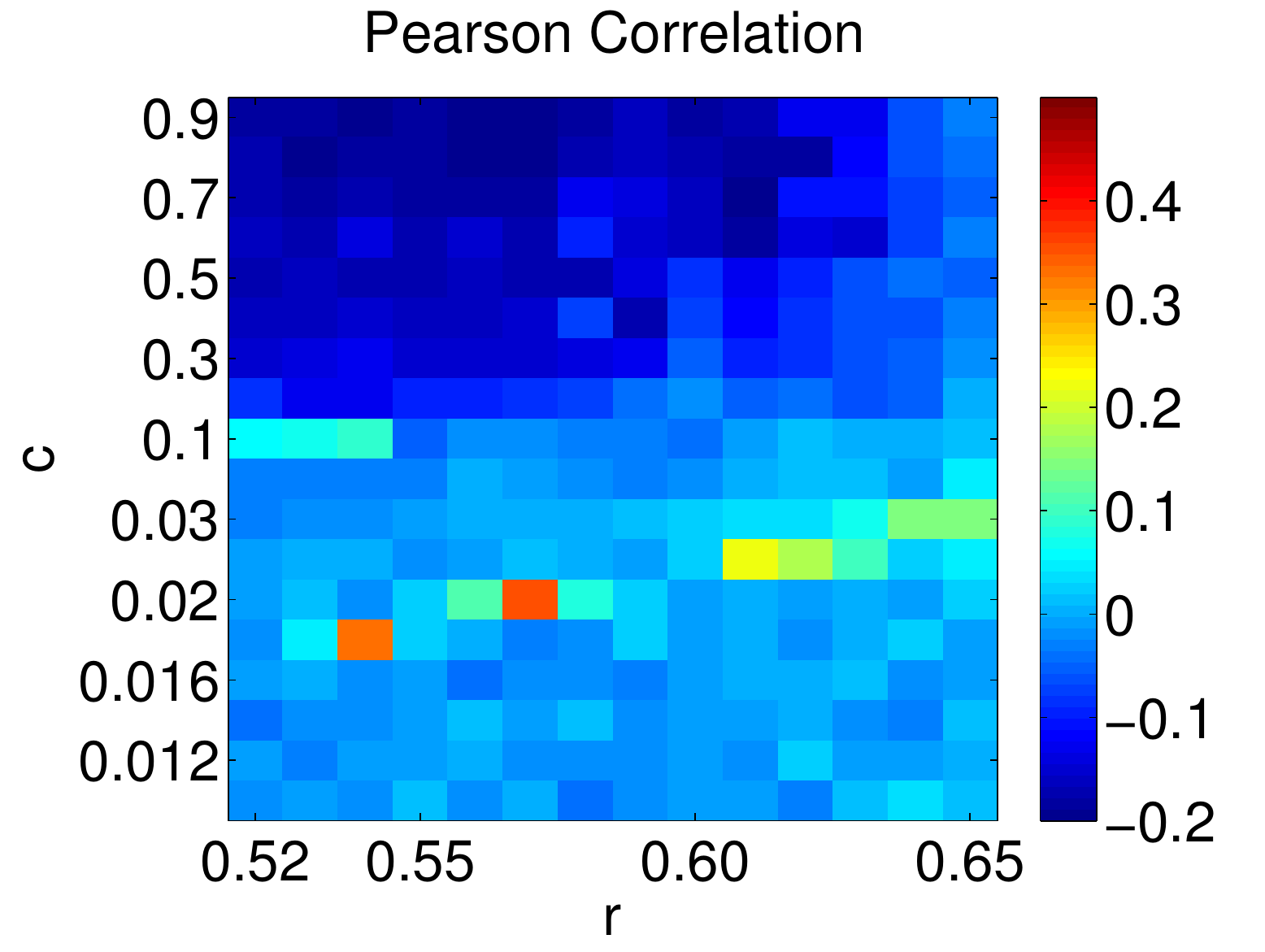}
{\bf(B)}\includegraphics[width=0.45\textwidth]{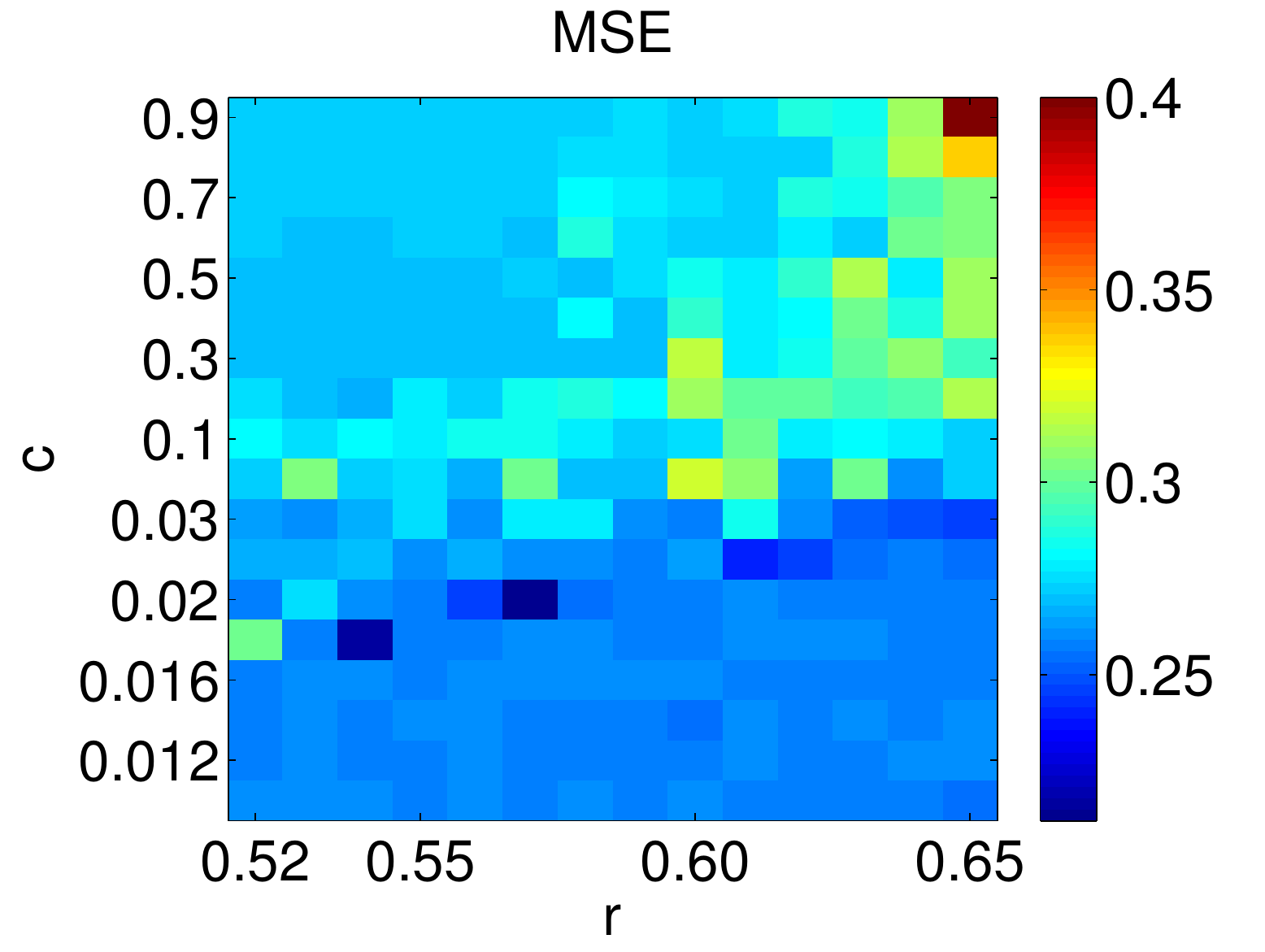}
\end{center}
\caption{
{\bf Pearson correlation coefficient} between experimental and simulated functional connectivity {\bf(A)} and means
squared error {\bf(B)} in the parameter space of the global coupling strength $c$ and 
the correlation threshold $r$. System
parameters as in Fig.~\ref{node_ts}.} 
\label{sim_exp}
\end{figure*}

In order to investigate complex collective network dynamics we study the global level of the network synchrony and its
variation, i.e. metastability as a function of the global coupling strength and coupling topology. We
simulated 7.5 min of the whole network dynamics to match experimental fMRI time series. Figure~\ref{sync} shows how
the level of network synchrony, calculated by time-averaged order parameter $\langle R(t)\rangle$ given in
Eq.~\ref{eq:eq4}, varies depending on the free model parameters $c$ and $r$. We see in Fig.~\ref{sync}(A) that the
global level of synchrony ranges from fully synchronized ($\langle R\rangle = 1$) to desynchronized ($\langle R\rangle =
0$) states, 
depending on the choice of $r$ and $c$. The resting-brain dynamics hardly show either extreme. Instead, they are 
characterized by metastable states. Hence, we calculate the standard deviation of the global order parameter 
as an indicator of this metastability, as shown in Fig.~\ref{sync}(B). 
For small $c$ and over all examined network configurations (all examined thresholds $r$), as well as for large $c$ and
higher values of the network density (low $r$), the networks show high synchronization indicated by red areas in
Fig.~\ref{sync}(A). This regime, however, corresponds to robust full synchronization, which does not allow large
variability. Accordingly, $R(t)$ does not show pronounced fluctuations. However, as shown in Fig.~\ref{sync}(B),
variations in synchrony display considerable level only for the narrow range of $c$ and $r$, as indicated by red color.
In this area of the parameter space, simulated network dynamics undergo permanent changes from synchronized to a less
synchronized state. At the same time, this area corresponds to the region where the agreement between the simulations
and the experiment is the best (as shown in  Fig.~\ref{sim_exp}). It can be seen that the simulations and the experiment
show best agreement within the region in the parameter space where 
the network shows considerable amount of synchrony ($\langle R\rangle \approx 0.5$) and metastability ($\sigma_R \approx 0.15$). Thus, we consider these values as the most relevant regions of the network synchrony and metastability.

To quantify the dynamic network's variations in synchrony, we calculated fast Fourier transform of time-averaged order
parameter $R(t)$. As shown in Fig.~\ref{fft_sync}(A), as the coupling strength between the nodes increases, the order
parameter starts to fluctuate in more periodic manner, indicating almost periodic network's transitions between
synchrony and asynchrony. The frequency of these transitions settles around 10 Hz, for $c \geq 0.02$. See
Fig.~\ref{fft_sync}(B).

To study the network properties that contribute to the dynamic behavior in the most relevant region, we have performed
graph theoretical analysis on the empirical network used as the coupling topology in the one of the simulation ($r =
0.57$). We have chosen this particular network configuration since the simulated dynamics based on this topology show
the best agreement with the experiment (see Fig.~\ref{sim_exp}). In Fig.~\ref{network_meas}, we show node-wise
network properties: degree, clustering coefficient and global efficiency. Clustering coefficient (CC) is used to
identify hubs in the network. CC values as high as 0.5 together with large degrees indicate that a node qualifies as 
hub \cite{NEW10}. Considering the nature of the interactions in the coupling topology we calculate the global node
efficiency (GE), which represents ``a measure of the overall capacity for parallel information transfer and integrated
processing'' \cite{BUL12}. We see that at this correlation threshold significant number of nodes show relatively high 
GE values. It has been already suggested that this network property play role in the metastable state in neural synchrony \cite{SHA10}.

\begin{figure}
\begin{center}
\includegraphics[width=0.45\textwidth]{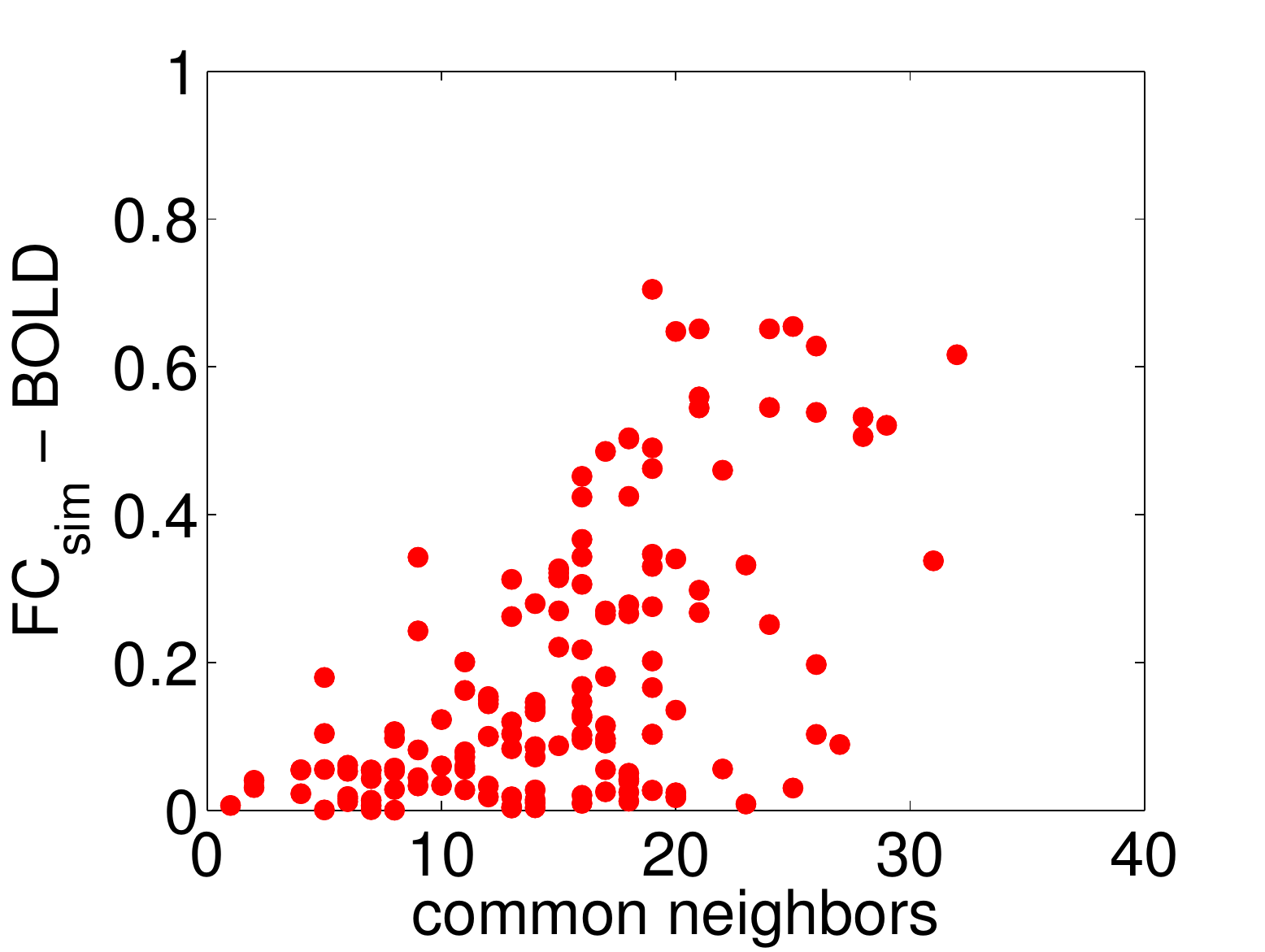}
\end{center}
\caption{
{\bf Functional connectivity (FC) between pairs of network nodes as a function of nearest neighbors.} Only nodes coupled
via indirect connections are represented. Data are shown for FC network simulated on the network topology at the
threshold $r = 0.57$ and for the coupling strength $c = 0.02$ (Pearson correlation $\rho = 0.51$ $p < 0.01$).
}
\label{indirect_fc}
\end{figure}

\begin{figure*}[!ht]
\begin{center}
\includegraphics[width=0.3\textwidth]{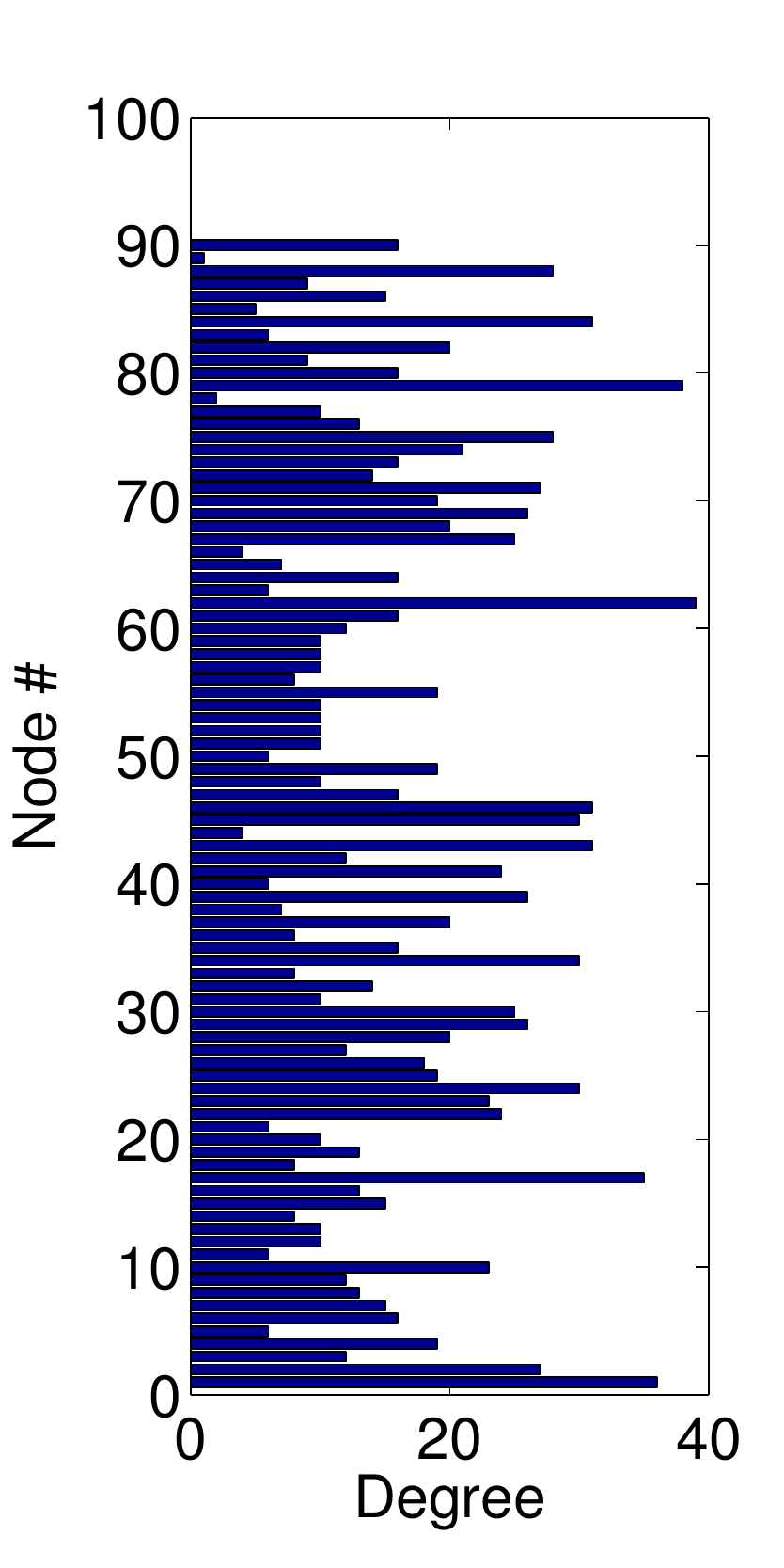}
\includegraphics[width=0.3\textwidth]{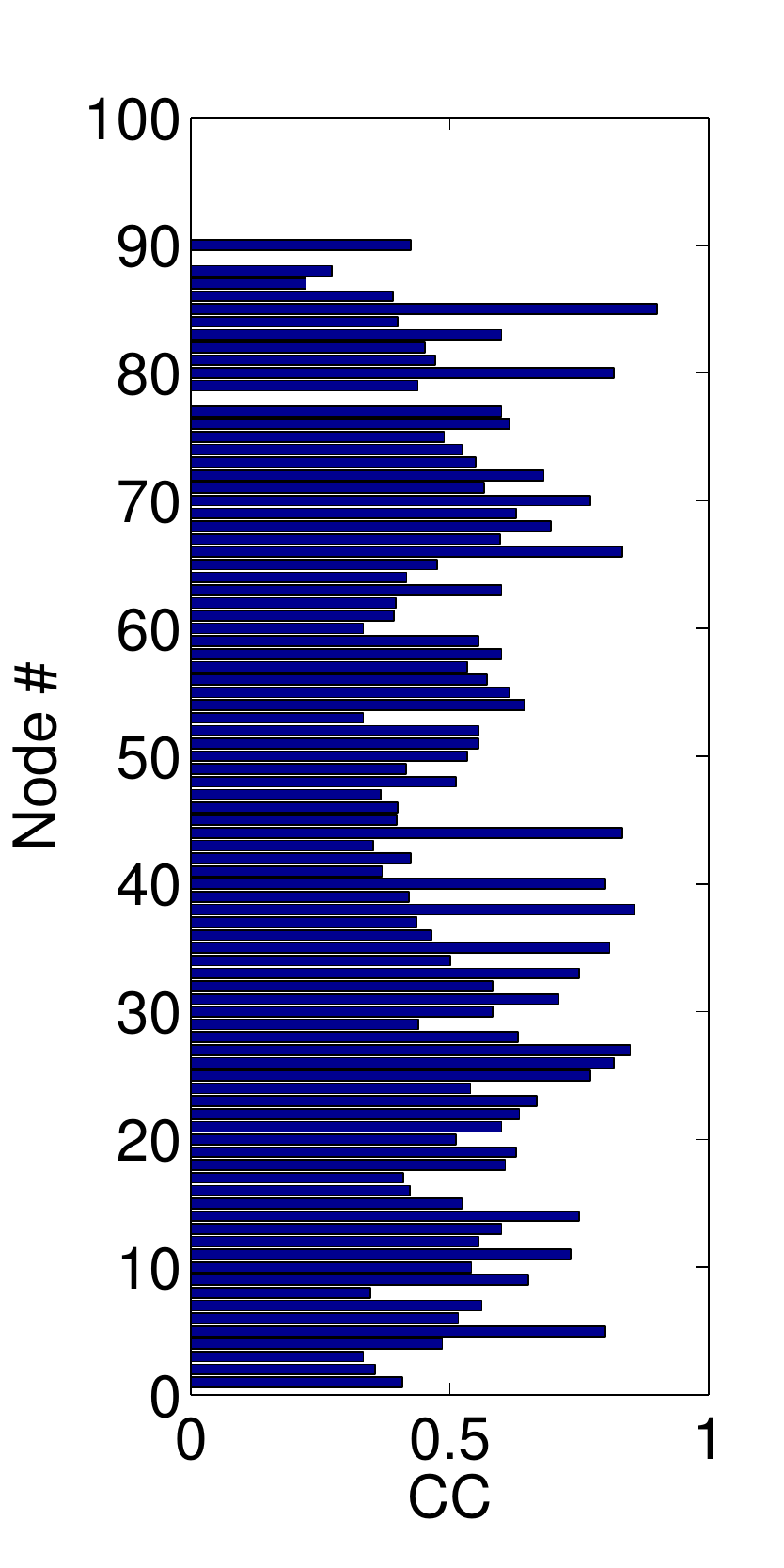}
\includegraphics[width=0.3\textwidth]{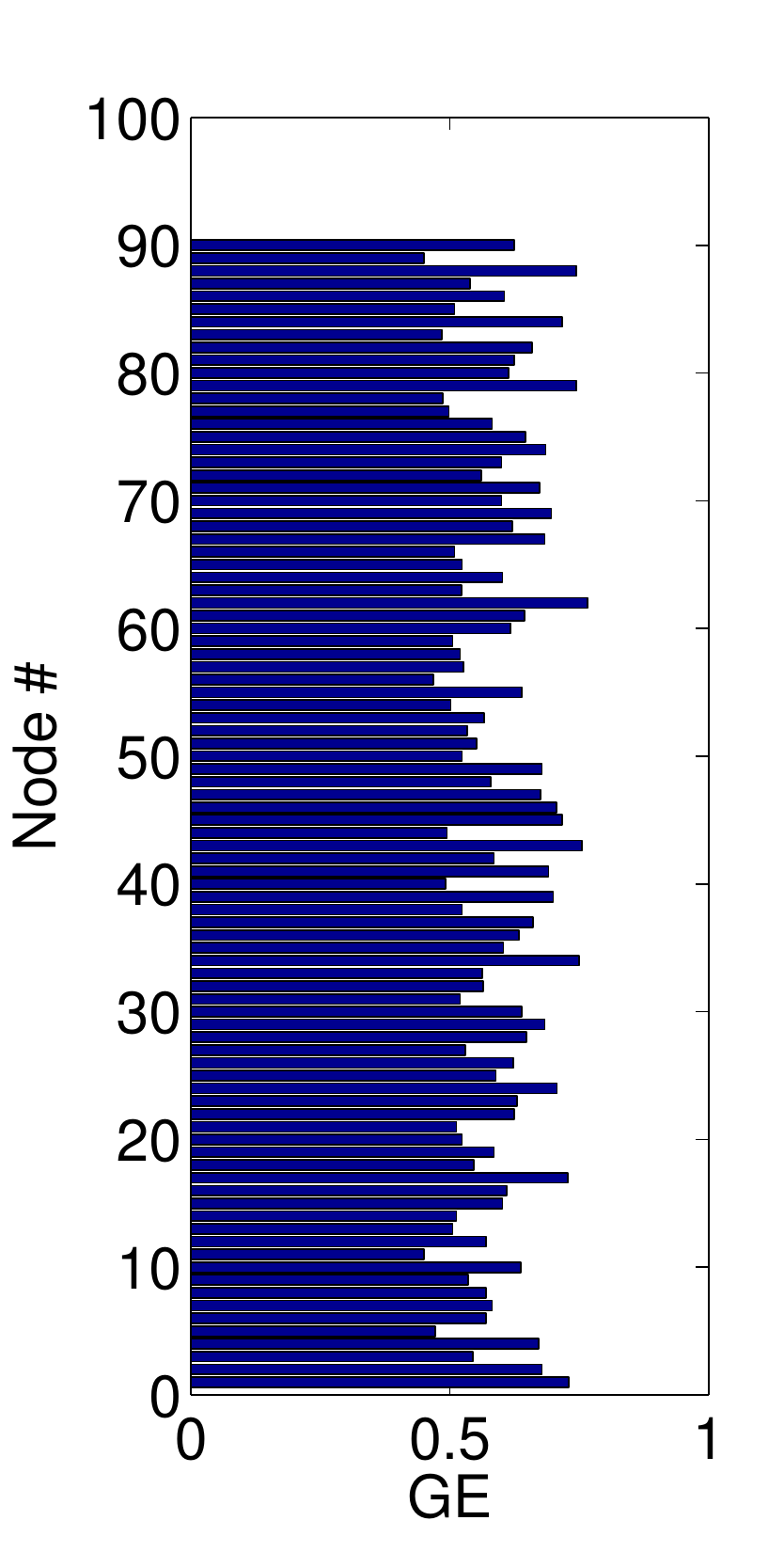}
\end{center}
\caption{
{\bf Node-wise complex network measures of the human brain functional 
interactions.} Degree, clustering coefficient (CC) and global efficiency 
(GE) of the network nodes of the empirical functional connectivity used in the simulations as coupling topology
(threshold $r = 0.57$).
}
\label{network_meas}
\end{figure*}

\section{Discussion}
In this work we have examined role of network configurations and synchronization dynamics in the emergence of resting-state functional correlations. Using numerical simulations of the neural network dynamics, based on the FHN neural models embedded into functionally realistic network structure and subject to noisy time-delayed interactions, we have investigated the network dynamics' transitions from synchronized to desynchronized state. We have found that the best agreement between the model and the experiment lie in the region where network dynamics maximizes synchrony and variations in synchrony. 

To understand how functional correlations between segregated brain areas arise from underlying functional interactions,
supported by the actual neural link between them, we use computational simulations and human brain imaging data. First,
we have extracted the network topology for our model using empirically obtained brain structural and functional
connectivities (see Fig.~\ref{acp_fc}). This takes into account changes in the network interactions similar to those
observed experimentally. We explore conditions that allow the dynamical properties of the brain networks  -- synchrony
and variations in synchrony -- to arise from different network configurations generating patterns of correlated activity
between the brain regions. By varying the network topologies in a range of experimentally realistic values \cite{BUL09,
WIJ10} and tuning the global coupling strength, we obtained the correlations patterns in the simulated BOLD time series
very similar to those found in the experiment. We show that flexible 
changes in the network dynamics -- fast periodic changes in the global level of synchrony (see Fig.~\ref{fft_sync}) -- produce functionally possible network dynamics. This shows how flexible changes in the large-scale networks dynamics could bee. However, we also show that this is the case only if both dynamical properties of the network -- synchrony and variations in synchrony -- are maximized (as indicated by green/yellow and red colors in Fig.~\ref{sync}(A) and (B), respectively). 

It is worth noting that in agreement with our previous study \cite{VUK14}, here we also observed a similar dependence of the functional correlations between indirectly coupled network nodes on the number of their overlapping neighbors (as
shown in Fig.~\ref{indirect_fc}). This important finding indicates that the result is robust with respect to the
generation of the network dynamics. As for Kuramoto-like phase oscillators, oscillating in $\gamma$
frequency range \cite{VUK14}, noise-driven FHN models, oscillating at around 15 Hz (i.e. in $\alpha$ band), display
similar dynamic behavior. The correlation patterns of low-frequency BOLD activity, revealing FC networks, highly agree
between the studies (see Fig. 5 in \cite{VUK14}) and also with the experiment. Moreover, these results are consistent
with some previous findings showing that resting-state FC networks arise from different types of local dynamics
\cite{HON07,DEC09,BRE10h,CAB11} and network topologies \cite{HON07,DEC09}. 
We also would like to point out a single value for the propagation velocity ($v = 7$ m/s) used in our study. As some
numerical studies have already shown, the system of the FHN neurons displays high sensitivity to the presence of the
time delays in the coupling term \cite{PAN12,OME13}. However, here we have not observed such a high sensitivity for the
biophysiologically valid velocity values (between 5 and 20 m/s) \cite{GHO08}. 

\section{Conclusion}
A main conclusion of our work is that fast flexible changes in neural network synchrony contribute to the emergence of correlated BOLD activity. By tuning the brain network topology and the dynamical interactions between the
regions (network nodes), we show that the model agreement with the experiment is the best for
a dynamical state that maximizes synchrony and variations in synchrony. These results support our
hypothesis that highly variable, metastable network dynamics may facilitate transitions between
network configurations.

\section{Acknowledgments}
This work was supported by BMBF (grant no. 01Q1001B) in the framework of BCCN 
Berlin (Project B7). We thank John-Dylan Haynes and his group for helpful 
discussions concerning the fMRI data analysis and Yasser Iturria-Medina for 
sharing the DTI data used in the study. 


\begin{thebibliography}{41}
\expandafter\ifx\csname natexlab\endcsname\relax\def\natexlab#1{#1}\fi
\expandafter\ifx\csname url\endcsname\relax
  \def\url#1{\texttt{#1}}\fi
\expandafter\ifx\csname urlprefix\endcsname\relax\def\urlprefix{URL }\fi

\bibitem[{Bassett et~al.(2014)Bassett, Wymbs, Porter, Mucha, and
  Grafton}]{BAS14}
Bassett, D.~S., Wymbs, N.~F., Porter, M.~A., Mucha, P.~J., Grafton, S.~T.,
  2014. Cross-linked structure of network evolution. Chaos: An
  Interdisciplinary Journal of Nonlinear Science 24~(1), 013112.

\bibitem[{Biswal et~al.(1995)Biswal, Yetkin, Haughton, and Hyde}]{BIS95}
Biswal, B., Yetkin, F.~Z., Haughton, V.~M., Hyde, J.~S., 1995. Functional
  connectivity in the motor cortex of resting human brain using echo-planar
  mri. Magnetic Resonance in Medicine 34~(4), 537--541.

\bibitem[{Bogacki and Shampine(1989)}]{BOG89}
Bogacki, P., Shampine, L.~F., 1989. A {3(2)} pair of runge - kutta formulas.
  Applied Mathematics Letters 2~(4), 321--325.

\bibitem[{Breakspear et~al.(2010)Breakspear, Heitmann, and
  Daffertshofer}]{BRE10h}
Breakspear, M., Heitmann, S., Daffertshofer, A., 2010. Generative models of
  cortical oscillations: neurobiological implications of the {Kuramoto} model.
  Frontiers in human neuroscience 4.

\bibitem[{Bullmore and Sporns(2009)}]{BUL09}
Bullmore, E.~T., Sporns, O., 2009. Complex brain networks: graph theoretical
  analysis of structural and functional systems. Nat. Rev. Neurosci. 10~(3),
  186--198.

\bibitem[{Bullmore and Sporns(2012)}]{BUL12}
Bullmore, E.~T., Sporns, O., 2012. The economy of brain network organization.
  Nature Reviews Neuroscience 13~(5), 336--349.

\bibitem[{Cabral et~al.(2013)Cabral, Fernandes, Van~Hartevelt, James, and
  Kringelbach}]{CAB13b}
Cabral, J., Fernandes, H.~M., Van~Hartevelt, T.~J., James, A.~C., Kringelbach,
  M. L. e.~a., 2013. Structural connectivity in schizophrenia and its impact on
  the dynamics of spontaneous functional networks. Chaos: An Interdisciplinary
  Journal of Nonlinear Science 23~(4), 046111.

\bibitem[{Cabral et~al.(2012)Cabral, Hugues, Kringelbach, and Deco}]{CAB12}
Cabral, J., Hugues, E., Kringelbach, M.~L., Deco, G., 2012. Modeling the
  outcome of structural disconnection on resting-state functional connectivity.
  NeuroImage 62, 1342--1353.

\bibitem[{Cabral et~al.(2011)Cabral, Hugues, Sporns, and Deco}]{CAB11}
Cabral, J., Hugues, E., Sporns, O., Deco, G., 2011. Role of local network
  oscillations in resting-state functional connectivity. NeuroImage 57~(1),
  130--139.

\bibitem[{Damoiseaux et~al.(2006)Damoiseaux, Rombouts, Barkhof, Scheltens,
  Stam, Smith, and Beckmann}]{DAM06}
Damoiseaux, J.~S., Rombouts, S. A. R.~B., Barkhof, F., Scheltens, P., Stam,
  C.~J., Smith, S.~M., Beckmann, C.~F., 2006. Consistent resting-state networks
  across healthy subjects. Proc. Natl. Acad. Sci. U.S.A. 103~(37),
  13848--13853.

\bibitem[{Deco and Jirsa(2012)}]{DEC12}
Deco, G., Jirsa, V.~K., 2012. Ongoing cortical activity at rest: criticality,
  multistability, and ghost attractors. The Journal of Neuroscience 32~(10),
  3366--3375.

\bibitem[{Deco et~al.(2011)Deco, Jirsa, and McIntosh}]{DEC11}
Deco, G., Jirsa, V.~K., McIntosh, A.~R., 2011. Emerging concepts for the
  dynamical organization of resting-state activity in the brain. Nature Reviews
  Neuroscience 12~(1), 43--56.

\bibitem[{Deco et~al.(2009)Deco, Jirsa, McIntosh, Sporns, and
  K{\"o}tter}]{DEC09}
Deco, G., Jirsa, V.~K., McIntosh, A.~R., Sporns, O., K{\"o}tter, R., 2009. {Key
  role of coupling, delay, and noise in resting brain fluctuations}. Proc.
  Natl. Acad. Sci. U.S.A. 106~(25), 10302--10307.

\bibitem[{Deco et~al.(2012)Deco, Senden, and Jirsa}]{DEC12a}
Deco, G., Senden, M., Jirsa, V.~K., 2012. How anatomy shapes dynamics: a
  semi-analytical study of the brain at rest by a simple spin model. Frontiers
  in computational neuroscience 6.

\bibitem[{Flunkert and Sch{\"o}ll(2009)}]{FLU09a}
Flunkert, V., Sch{\"o}ll, E., 2009. pydelay -- a python tool for solving delay
  differential equations. arXiv:0911.1633~[nlin.CD].

\bibitem[{Friston et~al.(1994)Friston, Holmes, Worsley, Poline, Frith, and
  Frackowiak}]{FRI94}
Friston, K., Holmes, A.~P., Worsley, K.~J., Poline, J.-P., Frith, C.~D.,
  Frackowiak, R., 1994. Statistical parametric maps in functional imaging: a
  general linear approach. Human brain mapping 2~(4), 189--210.

\bibitem[{Friston et~al.(2000)Friston, Mechelli, Turner, and Price}]{FRI00}
Friston, K., Mechelli, A., Turner, R., Price, C.~J., 2000. Nonlinear responses
  in {fMRI}: The balloon model, {Volterra} kernels, and other hemodynamics.
  NeuroImage 12~(4), 466--477.

\bibitem[{Ghosh et~al.(2008{\natexlab{a}})Ghosh, Rho, McIntosh, K{\"o}tter, and
  Jirsa}]{GHO08a}
Ghosh, A., Rho, Y., McIntosh, A.~R., K{\"o}tter, R., Jirsa, V.~K.,
  2008{\natexlab{a}}. Cortical network dynamics with time delays reveals
  functional connectivity in the resting brain. Cogn. Neurodyn. 2~(2),
  115--120.

\bibitem[{Ghosh et~al.(2008{\natexlab{b}})Ghosh, Rho, McIntosh, K{\"o}tter, and
  Jirsa}]{GHO08}
Ghosh, A., Rho, Y., McIntosh, A.~R., K{\"o}tter, R., Jirsa, V.~K.,
  2008{\natexlab{b}}. Noise during rest enables the exploration of the brain's
  dynamic repertoire. PLoS Comput Biol 4~(10), e1000196.

\bibitem[{Hellyer et~al.(2014)Hellyer, Shanahan, Scott, Wise, Sharp, and
  Leech}]{HEL14}
Hellyer, P.~J., Shanahan, M., Scott, G., Wise, R. J.~S., Sharp, D.~J., Leech,
  R., 2014. The control of global brain dynamics: Opposing actions of
  frontoparietal control and default mode networks on attention.
  J.~Neuroscience 34~(2), 451--461.

\bibitem[{Honey et~al.(2007)Honey, K{\"o}tter, Breakspear, and Sporns}]{HON07}
Honey, C.~J., K{\"o}tter, R., Breakspear, M., Sporns, O., 2007. {{N}etwork
  structure of cerebral cortex shapes functional connectivity on multiple time
  scales}. Proc. Natl. Acad. Sci. U.S.A. 104, 10240--10245.

\bibitem[{Honey et~al.(2009)Honey, Sporns, Cammoun, Gigandet, Thiran, Meuli,
  and Hagmann}]{HON09}
Honey, C.~J., Sporns, O., Cammoun, L., Gigandet, X., Thiran, J.~P., Meuli, R.,
  Hagmann, P., 2009. Predicting human resting-state functional connectivity
  from structural connectivity. Proceedings of the National Academy of Sciences
  of the United States of America 106~(6), 2035--2040.

\bibitem[{Iturria-Medina et~al.(2008)Iturria-Medina, Sotero,
  Canales-Rodr{\'\i}guez, Alem{\'a}n-G{\'o}mez, and Melie-Garc{\'\i}a}]{ITU08}
Iturria-Medina, Y., Sotero, R.~C., Canales-Rodr{\'\i}guez, E.~J.,
  Alem{\'a}n-G{\'o}mez, Y., Melie-Garc{\'\i}a, L., 2008. Studying the human
  brain anatomical network via diffusion-weighted mri and graph theory.
  Neuroimage 40~(3), 1064--1076.

\bibitem[{Izhikevich and Kuramoto(2006)}]{IZH06b}
Izhikevich, E.~M., Kuramoto, Y., 2006. Weakly coupled oscillators. Encyclopedia
  of mathematical physics 5, 448.

\bibitem[{Margulies et~al.(2007)Margulies, Kelly, Uddin, Biswal, Castellanos,
  and Milham}]{MAR07}
Margulies, D.~S., Kelly, A. M.~C., Uddin, L.~Q., Biswal, B.~B., Castellanos,
  F.~X., Milham, M.~P., 2007. Mapping the functional connectivity of anterior
  cingulate cortex. NeuroImage 37~(2), 579--588.

\bibitem[{Newman(2010)}]{NEW10}
Newman, M. E.~J., 2010. Networks: an introduction. Oxford University Press,
  Inc., New York.

\bibitem[{Nguyen and Hong(2011)}]{NGU11}
Nguyen, L.~H., Hong, K.-S., 2011. Synchronization of coupled chaotic
  fitzhugh--nagumo neurons via lyapunov functions. Mathematics and Computers in
  Simulation 82~(4), 590--603.

\bibitem[{Nicosia et~al.(2013)Nicosia, Valencia, Chavez, D{\'i}az-Guilera, and
  Latora}]{NIC13}
Nicosia, V., Valencia, M., Chavez, M., D{\'i}az-Guilera, A., Latora, V., 2013.
  Remote synchronization reveals network symmetries and functional modules.
  Phys. Rev. Lett. 110, 174102.

\bibitem[{Omelchenko et~al.(2013)Omelchenko, Omel'chenko, H{\"o}vel, and
  Sch{\"o}ll}]{OME13}
Omelchenko, I., Omel'chenko, O.~E., H{\"o}vel, P., Sch{\"o}ll, E., 2013. When
  nonlocal coupling between oscillators becomes stronger: patched synchrony or
  multichimera states. Phys. Rev. Lett. 110, 224101.

\bibitem[{Panchuk et~al.(2013)Panchuk, Rosin, H{\"o}vel, and
  Sch{\"o}ll}]{PAN12}
Panchuk, A., Rosin, D.~P., H{\"o}vel, P., Sch{\"o}ll, E., 2013. Synchronization
  of coupled neural oscillators with heterogeneous delays. Int. J. Bifurcation
  Chaos 23~(12), 1330039.

\bibitem[{Pikovsky et~al.(2001)Pikovsky, Rosenblum, and Kurths}]{PIK01}
Pikovsky, A., Rosenblum, M.~G., Kurths, J., 2001. Synchronization, A Universal
  Concept in Nonlinear Sciences. Cambridge University Press, Cambridge.

\bibitem[{Roy et~al.(2014)Roy, Sigala, Breakspear, McIntosh, and Jirsa}]{ROY14}
Roy, D., Sigala, R., Breakspear, M., McIntosh, A.~R., Jirsa, V.~K., 2014. Using
  the virtual brain to reveal the role of oscillations and plasticity in
  shaping brain's dynamical landscape. Brain Connectivity.

\bibitem[{Rubinov and Sporns(2010)}]{RUB10}
Rubinov, M., Sporns, O., 2010. Complex network measures of brain connectivity:
  uses and interpretations. NeuroImage 52~(3), 1059--1069.

\bibitem[{Seth et~al.(2013)Seth, Chorley, and Barnett}]{SET12}
Seth, A.~K., Chorley, P., Barnett, L.~C., 2013. Granger causality analysis of
  fmri bold signals is invariant to hemodynamic convolution but not
  downsampling. Neuroimage 65, 540--555.

\bibitem[{Shampine and Thompson(2001)}]{SHA01a}
Shampine, L.~F., Thompson, S., 2001. Solving {DDEs} in {Matlab}. Appl. Num.
  Math. 37~(4), 441--458.

\bibitem[{Shanahan(2010)}]{SHA10}
Shanahan, M., 2010. Metastable chimera states in community-structured
  oscillator networks. Chaos: An Interdisciplinary Journal of Nonlinear Science
  20~(1), 013108.

\bibitem[{Strogatz(2000)}]{STR00}
Strogatz, S.~H., 2000. From {K}uramoto to {C}rawford: exploring the onset of
  synchronization in populations of coupled oscillators. Physica D 143, 1--20.

\bibitem[{Tzourio-Mazoyer et~al.(2002)Tzourio-Mazoyer, Landeau, Papathanassiou,
  Crivello, Etard, Delcroix, Mazoyer, and Joliot}]{TZO02}
Tzourio-Mazoyer, N., Landeau, B., Papathanassiou, D., Crivello, F., Etard, O.,
  Delcroix, N., Mazoyer, B., Joliot, M., 2002. Automated anatomical labeling of
  activations in {SPM} using a macroscopic anatomical parcellation of the {MNI
  MRI} single-subject brain. Neuroimage 15~(1), 273--289.

\bibitem[{van Wijk et~al.(2010)van Wijk, Stam, and Daffertshofer}]{WIJ10}
van Wijk, B.~C., Stam, C.~J., Daffertshofer, A., 2010. Comparing brain networks
  of different size and connectivity density using graph theory. PLoS One
  5~(10), e13701.

\bibitem[{Vincent et~al.(2007)Vincent, Patel, Fox, Snyder, Baker, Van~Essen,
  Zempel, Snyder, Corbetta, and Raichle}]{VIN07a}
Vincent, J.~L., Patel, G.~H., Fox, M.~D., Snyder, A.~Z., Baker, J.~T.,
  Van~Essen, D.~C., Zempel, J.~M., Snyder, L.~H., Corbetta, M., Raichle, M.~E.,
  2007. Intrinsic functional architecture in the anaesthetized monkey brain.
  Nature 447~(7140), 83--86.

\bibitem[{Vuksanovi\'{c} and H{\"o}vel(2014)}]{VUK14}
Vuksanovi\'{c}, V., H{\"o}vel, P., 2014. Functional connectivity of distant
  cortical regions: Role of remote synchronization and symmetry in
  interactions. NeuroImage 97, 1--8.

\end{thebibliography}




\clearpage
\begin{table*}[!ht]
\caption{
\bf{Cortical and sub-cortical regions as defined in the automated 
anatomic 
labelling (AAL) template image.}}
\centering
\vspace{0.5cm}
\resizebox{6cm}{!}{
\begin{tabular}{|l|l|r|}
    \hline
    Index  & Anatomical Description & Label \\
    \hline
    1 & Precentral & PRE \\
    2 & Frontal Sup  & F1 \\
    3 & Frontal Sup Orb & F10 \\
     4 & Frontal Mid &  F2 \\
     5 & Frontal Mid Orb & F20 \\
     6 & Frontal Inf Oper & F30P\\
     7 & Frontal Inf Tri & F3T \\
     8 & Frontal Inf Orb & F30\\
     9 & Rolandic Oper & RO \\
     10 & Supp Motor Area & SMA \\
     11 & Olflactory & OC\\
     12 & Frontal Sup Medial & F1M\\
     13 & Frontal Mid Orb & SMG\\
     14 & Gyrus Rectus & GR \\
     15 & Insula & IN \\
     16 & Cingulum Ant & ACIN\\
     17 & Cingulum Mid & MCIN\\
     18 & Cingulum Post &  PCIN\\
     19 & Hippocampus & HIP\\
     20 & ParaHippocampal & PHIP\\
     21 & Amygdala & AMYG\\
     22 & Calcarine & V1\\
     23 & Cuneus & Q\\
     24 & Lingual & LING\\
     25 & Occipital Sup & O1\\
     26 & Occipital Mid & O2\\
     27 & Occipital Inf & O3\\
     28 & Fusiform & FUSI\\
     29 & Postcentral & POST\\
     30 & Parietal Sup & P1\\
     31 & Parietal Inf & P2\\
     32 & Supra Marginal Gyrus & SMG \\
     33 & Angular & AG\\
     34 & Precuneus &  PQ\\
     35 & Paracentral Lobule & PCL\\
     36 & Caudate & CAM\\
     37 & Putamen & PUT\\
     38 & Pallidum & PAL\\
     39 & Thalamus & THA\\
     40 & Heschi & HES\\
     41 & Temporal Sup & T1 \\
     42 & Temporal Pole sup & T1P\\
     43 & Temporal Mid & T2\\
     44 & Temporal Pole Mid & T2P\\
     45 & Temporal Inf & T3 \\
     \hline
  \end{tabular}
  }
\begin{flushleft}Indexes from 1-45 indicate order in which regions in the right hemisphere are arranged 
in all connectivity/adjacency matrices presented throughout the paper. For the respective regions in the left
hemisphere, the index needs to be shifted by 45.
\end{flushleft}
\label{tab:AAL}
\end{table*}

\end{document}